\documentclass[aps,prl,onecolumn,reprint,floatfix, superscriptaddress]{revtex4-2}

\usepackage{subfigure}
\usepackage{psfrag,graphicx}
\usepackage{pict2e}
\usepackage{dcolumn}
\usepackage{amsmath,amssymb,braket}
\usepackage{bm}
\usepackage{latexsym}
\usepackage{epstopdf}
\usepackage[english]{babel}
\usepackage[utf8]{inputenc}
\UseRawInputEncoding
\usepackage{amsfonts}
\usepackage{bm}
\usepackage{natbib}
\usepackage{bbold}
\usepackage{enumerate}
\usepackage{xcolor}
\definecolor{Cerulean}{rgb}{0.,0.59,0.835}
\definecolor{RubineRed}{rgb}{0.61,0.07,0.12}
\definecolor{myblue}{rgb}{0.2,0.2,0.8}
\usepackage[colorlinks=true,citecolor=myblue,linkcolor=RubineRed,urlcolor=Cerulean]{hyperref}

\newcommand\jj{\mathbf{j}}

\newcommand\kk{\mathbf{k}}

\newcommand\stx[1]{_\text{#1}}
\newcommand\rr{\mathbf{r}}

\newcommand\hc{\text{h.c.}}

\newcommand{\pa}[1]{\left( #1\right)}
\newcommand{\co}[1]{\left[ #1\right]}
\newcommand{\abs}[1]{\left| #1\right|}
\newcommand{\ev}[1]{\left\langle #1\right\rangle}

\newcommand{\ham}[1]{\hat H_{\text{#1}}}

\def\ii{{\textbf i}}
\def\jj{{\textbf j}}

\def\kk{{\textbf k}}

\def\rr{{\textbf r}}
\def\ss{{\textbf s}}

\begin{document}

\title{Optical lattice quantum simulator of dynamics beyond Born-Oppenheimer}
\author{Javier Arg\"uello-Luengo}
 \email{javier.arguello.luengo@upc.edu}
\affiliation{Departament de F\'isica, Universitat Polit\`ecnica de Catalunya, Campus Nord B4-B5, 08034 Barcelona, Spain}
\author{Alejandro~Gonz\'{a}lez-Tudela}
\email{a.gonzalez.tudela@csic.es}
\affiliation{Institute of Fundamental Physics IFF-CSIC, Calle Serrano 113b, 28006 Madrid, Spain.}
\author{J. Ignacio Cirac}
\affiliation{Max-Planck-Institut fÜr Quantenoptik, Hans-Kopfermann-Str. 1, 85748 Garching, Germany}
\affiliation{Munich Center for Quantum Science and Technology (MCQST), Schellingstr. 4, 80799 M\"unchen, Germany}

\begin{abstract}
 Here, we propose a platform based on ultra-cold fermionic molecules trapped in optical lattices to simulate nonadiabatic effects, as they appear in certain molecular dynamical problems. The idea consists of a judicious choice of two rotational states as the simulated electronic or nuclear degrees of freedom, in which their dipolar interactions induce the required attractive or repulsive interactions between them. 
 We benchmark our proposal by studying the scattering of an electron or a proton against a hydrogen atom, showing the effect of electronic exchange and inelastic ionization as the mass ratio between the simulated nuclei and electrons --a tunable experimental parameter in our simulator-- becomes comparable. These benchmarks illustrate how the simulator can qualitatively emulate phenomena like those appearing in molecular dynamical problems even if the simulated interaction occurs in two-dimensions with a dipolar scaling. Beyond the molecular implementation proposed here, our proposal can be readily extrapolated to other atomic platforms, e.g., based on fermionic Rydberg atoms.
\end{abstract}
\maketitle

Cold atomic systems in optical lattices have been one of the leading platforms to simulate quantum many-body physics over the past two decades~\cite{Lewenstein2007, Gross2017}. Their combination of exceptional control~\cite{Bloch2012} and detection techniques~\cite{Gross_2021} has enabled the study of relevant many-body phenomena in condensed matter~\cite{Hart_2015, Boll_2016, Cheuk_2016, Mazurenko_2017, Xu_2023, Shao_2024,Chiu_2019, Bourgund_2023,Sompet_2022,Hirthe_2023, Hartke_2023,Schreiber_2015, Nichols_2019, Guardado-Sanchez_2020, Scherg_2021} or lattice high-energy physics problems~\cite{Banuls_2020, Aidelsburger2022, DiMeglio_2024}. Recently, a new avenue has opened in optical lattice setups with proposals to simulate few-body problems~\cite{arguello-luengoAnalogue2019,arguello-luengoQuantum2020,arguello-luengoEngineering2021,malz2023} similar to those appearing in quantum chemistry, but with different interaction scalings and dimensionalities. Compared to other proposals for trapped ion simulators~\cite{MaDonnel2021,Navickas2024,Valahu2023,Olaya2024,haAnalog2024}, these optical lattice simulators natively encode the electronic degrees of freedom in the fermionic atoms hopping along the lattice which, combined with quantum gas microscopes, enable unique capabilities for the detection of electronic correlations. However, these atomic proposals have several limitations. First, they focus on the exploration of ground-state physics rather than dynamical properties~\cite{arguello-luengoAnalogue2019,arguello-luengoQuantum2020,arguello-luengoEngineering2021,malz2023}, where these analog simulators have their true power. Second, to emulate the required interactions, they either extend the range of local on-site interactions through complex laser-assisted processes~\cite{arguello-luengoAnalogue2019,arguello-luengoQuantum2020,arguello-luengoEngineering2021} or by harnessing Rydberg interactions, which are limited by spontaneous emission~\cite{malz2023}. Last, and most importantly, the nuclei are represented by classical fixed potentials, and thus cannot capture nonadiabatic dynamical effects. 

\begin{figure}[tb]
    \centering
    \includegraphics[width = \columnwidth]{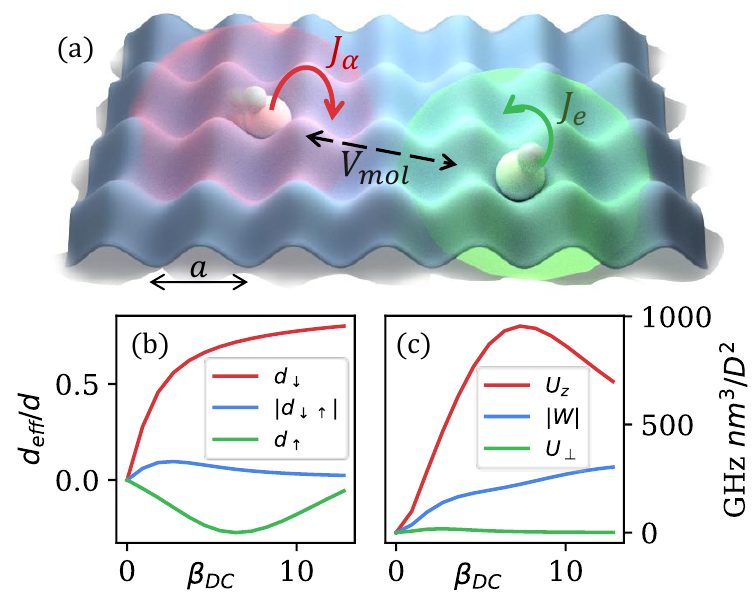}
\caption{(a) Scheme of the simulator, where molecules in two different rotational levels $\ket{\downarrow}$ and $\ket{\uparrow}$ (represented in red and green along the text) tunnel in an optical lattice with rates $J_{\alpha}$ and $J_e$, playing the role of nuclei and electrons, respectively. The dipolar interactions between these molecular levels satisfy the desired attractive and repulsive nature of the simulated long-range forces. (b) Effective dipolar moment of rotational levels $\ket{\downarrow}=\ket{00}$ and $\ket{\uparrow}=\ket{20}$, as well as the overlap $d_{\downarrow\uparrow}=\bra{\downarrow} \hat d_0 \ket{\uparrow}$ for different values of external DC electric field, $\beta\stx{DC}\equiv E_0 d/B_N$ (see main text). (c) Components of the resulting spin Hamiltonian of Eq.~\eqref{eq:Vmol} for increasing values of the electric field.\label{fig:scheme}}
\end{figure}

In this Letter, we propose a new strategy that overcomes the aforementioned limitations by harnessing recent advances for molecules trapped in an optical lattice. By using two dressed rotational levels, the system incorporates both the electronic and nuclear degrees of freedom as dynamical variables, thus providing a platform to explore a broader class of quantum effects beyond the constraints of the Born-Oppenheimer approximation. Importantly, we show that by adding an external electric field, there exists a regime where the dipolar interactions between these molecular states can emulate the repulsive nucleus-nucleus and electron-electron interactions, while at the same time yielding attractive nucleus-electron forces between the different levels. To benchmark the simulator, we illustrate how it can emulate the scattering of an electron or proton (H$^+$) against a hydrogen atom ($H$), showing how to prepare the initial scattering wavepackets and extract the resulting scattering cross-section. Importantly, we show that the proposed simulator can access the nonadiabatic regime due to the ability to tune the effective nucleus vs.~electron mass ratio experimentally.

\noindent \emph{Complete quantum-chemistry Hamiltonian.-} The simulation of molecular problems starts by choosing an appropriate basis to write the second-quantized quantum chemistry Hamiltonian. A natural choice for cold-atom simulators is the grid discretized basis, where one chooses one fermionic mode per point of space and degree of freedom. Compared to Refs.~\cite{arguello-luengoAnalogue2019,arguello-luengoQuantum2020,arguello-luengoEngineering2021,malz2023}, focused on the electronic system, here we also need to include additional operators to account for the nuclear degrees of freedom. In contrast to other digital approaches based on projected electronic states~\cite{szabo2012modern}, this natural mapping of an individual atom to each simulated electron or nucleus results in simple two-body density interactions, and the final complete chemistry Hamiltonian we aim to simulate reads:

\begin{equation}
\hspace*{-0.2in}
    \begin{split}
\label{eq:Htarget}
&\hat{H} =-\sum_\alpha J_{\alpha}\sum_{\langle \ii, \jj \rangle} \hat{c}^{\dagger}_{\alpha\ii} \hat{c}_{\alpha\jj}
- J_e\sum_{\sigma,\langle \ii, \jj \rangle} \hat{f}^{\dagger}_{\sigma\ii} \hat{f}_{\sigma\jj}  \\
& + \sum_{\ii,\jj} V_{\ii,\jj} \co{ \sum_{\sigma,\sigma'}\hat n^f_{\sigma\ii} \hat n^f_{\sigma'\jj} + \sum_{\alpha\neq \alpha'} Z_{\alpha} Z_{\alpha'} \hat n^c_{\alpha\ii} \hat n^c_{\alpha'\jj} - \sum_{\sigma,\alpha} Z_\alpha \hat n^c_{\alpha\ii} \hat n^f_{\sigma\jj}} 
\end{split}
\end{equation}

where $\hat{c}^{(\dagger)}_{\alpha\ii}$ is the annihilation (creation) operator of nucleus $\alpha$ at site $\ii$, and $\hat{f}^{(\dagger)}_{\sigma\ii}$ are the ones of fermions with spin $\sigma$. $\hat n^f_{\sigma\ii}=\hat{f}^{\dagger}_{\sigma\ii} \hat{f}_{\sigma\ii}$ and $\hat n^c_{\alpha\ii}=\hat{c}^{\dagger}_{\alpha\ii} \hat{c}_{\alpha\ii}$ are the electronic and nuclear number operators, respectively. The first line in Eq.~\eqref{eq:Htarget} represents the kinetic terms, where $J_{\alpha}$ and $J_e$ are the hopping amplitudes for nucleus $\alpha$ and the electronic degrees of freedom, respectively. The second line of the Hamiltonian,  $\hat V$, indicates the extended interaction among electrons and nuclei, which are weighted by their charge number, $Z_\alpha$. 
Based on this, in the following, we propose a two-dimensional model that exhibits an extended force $V_{\ii,\jj}=V\pa{|\ii-\jj|}$ with the correct signs, so that the simulated nuclei and electrons repel themselves, but are attracted to each other. 
Although quantitatively different from real scattering events, these simplifications preserve the nonadiabatic effects and correlated electron-nucleus states that we will later explore.

\noindent \emph{Simulator setup.-} A general scheme of the setup is depicted in Fig.~\ref{fig:scheme}(a): individual molecules are trapped at the minima of a two-dimensional optical lattice with spacing $a$ and $N_{x,y}$ sites per side~\cite{christakisProbing2023,yan13a,chotiaLongLived2012}. An important aspect is that, for $m$ different types of nuclei involved in the simulation, one needs $m+2$ internal levels to encode the nuclear and electronic degrees of freedom of the molecular Hamiltonian of Eq.~\eqref{eq:Htarget}. For now, we focus on one electronic spin component and assume that all nuclei are of the same type, so that nuclear and electronic creation operators, $\hat c$ and $\hat f$, only require access to two different rotational levels.  

A crucial aspect of this simulator is the choice of rotational levels, ensuring repulsion when molecules are in the same state and attraction when in different states. For that, one can choose $^1\Sigma$ polar molecules~\cite{brownRotational2003}, where there are no unpaired electrons, and the electronic wavefunction is invariant under all symmetry transformations, as there is neither orbital nor spin angular momentum~\cite{wall2015Quantum}. These molecules have well-isolated internal rotational levels $\ket{N,M}$~\cite{micheliToolbox2006}, in which the first index denotes the rotational angular momentum quantum number associated to $\textbf{N}^2$, and the second one is its projection along the quantization axis. In the absence of fields, these energy levels are $(2N+1)$-fold degenerate, since the molecules are described by an effective rigid-rotor Hamiltonian $\ham{rot}=B_N \textbf{N}^2 $, where $B_N$ is the rotational constant. To break this degeneracy, we add an external DC field, $\ham{DC}=-\textbf{d}\cdot \textbf{E}\stx{DC}=-\hat d_z E_0$, that renormalizes its effective dipole moment. This field is aligned along the z-axis to preserve the isotropy of the simulation over the optical lattice situated in the XY plane. As a result, the lowest rotational state, $\ket{00}\equiv \ket{\downarrow}$, which would otherwise be rotationally symmetric in the absence of fields, acquires an effective positive dipolar moment $d_{\downarrow}=\bra{\downarrow}\hat d_0 \ket{\downarrow}$, as shown in Fig.~\ref{fig:scheme}(b). Other states, like $\ket{20}\equiv \ket{\uparrow}$, antialign with the field at intermediate values $\beta\stx{DC}\equiv E_0 d/B_N\sim 1$, as also shown in Fig.~\ref{fig:scheme}(b), which enables inducing an attractive interaction between molecules when they are in these different states. This can be explicitly shown by projecting the dipolar interaction Hamiltonian onto the reduced subspace formed by these two states~\cite{wall2015Quantum}:
\begin{equation}
\begin{split}
    \label{eq:Vmol}
    \hat V\stx{mol}=\sum _{\ii,\jj} \frac{1}{|\rr_\ii-\rr_\jj|^3}&\Big{[}\frac{U_\bot}{2}\pa{\hat S_\ii^+ \hat S_\jj^- + \hc}\\
   & +U_z \hat S_\ii^z \hat S_\jj^z + W\pa{\hat S_\ii^z + \hat S_\jj^z}\Big{]}\,,
\end{split}
\end{equation}
where $U_\perp \equiv 2 d_{\downarrow \uparrow}^2\,,$ $U_z \equiv ( d_{\uparrow}-d_{\downarrow})^2\,,$ and $W \equiv \pa{d_{\uparrow}^2 - d_{\downarrow}^2}/2\,$~\cite{gorshkovTunable2011, micheliCold2007, micheliToolbox2006, wallHyperfine2015, wall2015Quantum}. Here, $d_\sigma=\bra{\sigma}\hat d_0 \ket{\sigma}$, and $d_{\downarrow\uparrow}=\bra{\uparrow}\hat d_0 \ket{\downarrow}$ are the projections of the dipole operator in the rotational subspace $\{\ket{\downarrow},\ket{\uparrow}\}$ where the spin operators $\hat S_i$ are defined. In Fig.~\ref{fig:scheme}(c), we plot these parameters ($ U_z, W, U_\perp$) for increasing values of the electric field. There, we observe how for $\beta\stx{DC}\approx8$, state-of-the-art lattice spacings $a\sim 500$ nm, and permanent dipole moments $d\sim 1$ Debye (as is the case for molecules such as KRb or NaRb~\cite{yanResonant2020,christakisProbing2023}), one obtains a nearest-neighbor value $V_0=U_z/a^3\sim 10$ kHz, which is in the order of the tunneling time. In addition, we find that the detrimental mixing between rotational states is reduced to $U_\bot / U_z \sim 1 \%$. The latter term, $W$, corrects the strength of these interactions but does not change their scaling~\cite{EM}. As a result, one is left to this order with an effective Hamiltonian that is diagonal in the $z-$basis where molecules on the same state experience a repulsive interaction due to the alignment of their charges, while different states attract each other. In both cases, we will consider an interaction of the form $V\stx{mol}(r)=V_0a^3/\pa{r_0^3 +r^3}$, which captures both the $1/r^3$ scaling of dipolar interactions and the renormalization at short distances, $r_0\sim a$, associated with on-site interactions~\footnote{While the effective Hamiltonian holds valid for inter-site interactions, dipolar interactions will dominate over the electric field for separations $r<r_d\equiv (d^2/B_N)^3 \sim 10$ nm. As $r_d<a$, this will only affect on-site interactions, which motivates our choice of effective potential $V\stx{mol}$~\cite{yanResonant2020,janssenCold2012}. Microwave shielding to repulsive resonant dipolar interactions could potentially be used to prevent molecules from getting closer to each other~\cite{karmanResonant2022,yanResonant2020}.}. Although it differs from the $1/r$ Coulomb potential that characterizes quantum chemistry, the induced two-dimensional extended forces encode the correct attractive and repulsive character of the interactions among electrons and nuclei in Eq.~\eqref{eq:Htarget}. Relevantly, the presence of nuclear motion in the simulation allows one to access regimes where nonadiabatic effects appear. This is especially relevant for chemical reactions and scattering dynamics that are influenced by chirality~\cite{lombardiChirality2018} or the presence of exceptional points~\cite{domckeConical2004,meadDetermination1979,wuPrediction1993,althorpeObservation2002}, where numerical methods based on the Born-Oppenheimer approximation are compromised. Given the difficulty to numerically tackle few-body problems in those regimes, in the following we will explore some simplified scenarios that we can numerically study in detail to validate and characterize the simulator, guiding the initial configurations for the experimental exploration of this field.

\begin{figure}[!tbp]
    \centering
    \includegraphics[width = \columnwidth]{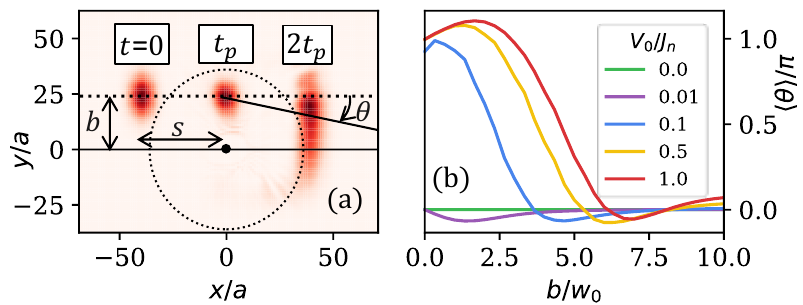}
\caption{(a) Temporal evolution of a 2D wavepacket of width $w_0/a=5.66$ that propagates in the right direction (dotted line) with carrier wavevector $k_0a=\pi/2$ for a lattice with $140\times120$ sites, and $V_0/J_n=0.4$. It has a vertical impact parameter $b/w_0=1.06$, and the initial horizontal separation is $s/a=40$ sites. The Mie potential that emulates the nucleus is centered at the origin and the dotted circumference indicates a radius $r_M/a=36$. (b) Expected scattering angle as a function of the impact parameters for Mie potentials of different strengths.}
\label{fig:scattering}
\end{figure}

\noindent \emph{Simulating single-particle scattering.-} We first start benchmarking the simulator by studying its potential to emulate classical scattering processes. For that, we consider a single dynamical \emph{projectile} (a proton or electron, represented by the corresponding rotational level of a molecule) against a fixed classical \emph{target} described by a Mie potential of the form $V\stx{Mie}(r,r_M)=V_0 \co{(r/r_M)^{-2}-2(r/r_M)^{-1}}$, which is repulsive at short distances, $r\ll r_M$, and exhibits an attractive quadratic local minimum at distance $r_M$. In an experiment, this fixed potential can be easily engineered with an optical potential defined, e.g., by an intensity phase mask~\cite{rubio-abadalManyBody2019}. 

The projectile is prepared as a gaussian wavepacket $\psi_{w_0,k_0}(\rr)$ whose initial width $w_0$ extends over several sites of the lattice and carries initial momentum $k_0$ along the horizontal axis (see Ref~\cite{EM} for further numerical details). It has an initial horizontal separation $s$ from the central potential, which is reached maximally at time $t_p=s/(2Ja)$. Using cold atoms, the amplitude of this wavepacket can be prepared with the expansion of a localized state~\cite{fukuharaQuantum2013, karskiQuantum2009,youngTweezerprogrammable2022}, and the correct phase can be spatially imprinted using light-modulation~\cite{lauberAtomic2011,fabreRealization2011} or the reflection with barrier potentials~\cite{suColdAtom2024}. In two dimensions, this moving wavepacket would however suffer from an undesired dispersion along the vertical axis, which results into an additional width $w(t)$ and quadratic phase $\varphi(\rr,t)$ that increase as the projectile propagates~\cite{EM}. To reduce this distortion at impact time $t_p$, here we propose a different strategy by initially preparing the wavepacket $\psi\stx{in}(\rr,0)=\psi_{w(t_p),k_0}(\rr)\exp\co{-i\varphi(\rr,t_p)}$, which dynamically compensates for these effects so that the target is reached by an undistorted gaussian state.

In Fig.~\ref{fig:scattering}(a) we superpose the probability density, $\abs{\psi\stx{in}(\rr,t)}^2$, of three different instants in the scattering of this self-focusing wavepacket. For an \emph{impact factor} (vertical separation, $b$) comparable to the length of the attractive region of the potential, $b\approx r_M$, one encounters the \emph{glory impact factor} where the wavepacket maximally bends toward the center of the potential, as we can appreciate in the final frame. In Fig.~\ref{fig:scattering}(b), we calculate the average scattering angle $\ev{\theta}$ away from the incoming direction for increasing values of impact parameters. For direct collisions $(b=0)$ the wavepacket is scattered backward due to the repulsive central region of the potential [$\ev{\theta}\approx \pi$]. As the scattering strength of the nuclear potential $V_0/J_e$ increases, we observe that the most negative scattered angle appears for larger glory impact parameters.

\begin{figure}[!tbp]
    \centering
    \includegraphics[width = 0.95\columnwidth]{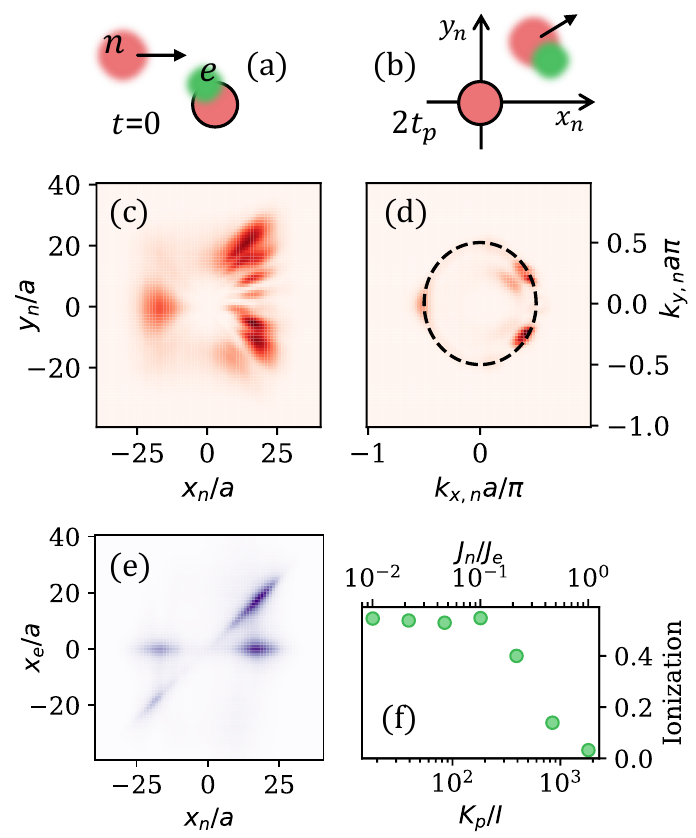}
\caption{(a,b) Scheme for nuclear scattering against hydrogen at times $0$ and $2t_p$, respectively. Blurry contours indicate that the incoming nucleus [n] and target electron [e] are treated quantum-mechanically, while the contoured nucleus is fixed. Nuclear probability density of the scattered nucleus in real (c) and momentum space (d). Dashed line indicates momentum $k_0a=\pi/2$. (e) Spatial correlations between the scattered nucleus and the ejected electron along the axis parallel to the incoming nucleus. (f) Electronic excitation rate as a function of the tunneling rate of the incoming nuclear wavepacket. \emph{Parameters:} $N_{x,y}=80$, $V_0/J_e=0.4$, $w_0/a=4.25$, $r_0/a=1.1$, $J_n/J_e=0.022$, $k_0a=\pi/2$, $b/a=3$.}
\label{fig:nucel}
\end{figure}

 \noindent \emph{Simulating electron exchange.-} We now study the electron exchange when a proton impacts a hydrogen atom, as schematized in Fig.~\ref{fig:nucel}(a,b). For the numerical benchmark presented here, we consider a dynamical proton scattering against a dynamical electron bounded to a fixed nuclear potential $V\stx{mol}(r)$, which can be fixed optically in the experiment. One should note that the ratio between effective incoming kinetic energy $K_p=2J_n(1-\cos{k_0 a})$ and the ionization energy of the target hydrogen, $I$, can be controlled through the carrier wavevector $k_0$, or the nuclear dynamics $J_n<J_e$. To minimize diffusive processes along the projectile direction, we choose the linear region of the dispersion relation $k_0a=\pi/2$, which still allows us to tune $K_p=2J_n$, through the nuclear tunneling rate $J_n$. This tunability can benefit from recent experiments with state-dependent lattices~\cite{bauseTuneOut2020,yeQuantum2008,guanMagic2021}. As the electron in the target hydrogen feels the attraction of the incoming proton, it can be either released from its parent nucleus or become bounded to the propagating projectile after the scattering event. In Fig.~\ref{fig:nucel}(c) we illustrate the scattered states of the projectile at time $2t_p$, for the impact parameter $b/a=3$ and $J_n/J_e=0.022$, where we observe the presence of diffraction fringes in the final state due to interactions with the target. In momentum space [Fig.~\ref{fig:nucel}(d)], we observe a larger emission in the forward direction. Dashed circle indicates the initial carrier momentum $k_0$, which highlights the reduced kinetic energy in the projectile due to the inelastic energy transfer to the electron in the target hydrogen. In Fig.~\ref{fig:nucel}(e), we show the horizontal spatial correlations between the scattered proton and target electron, which confirms that the electron remains bound to the parent nucleus (horizontal correlation), or associates with the incoming projectile in an exchange process (diagonal correlation).

In a Born-Oppenheimer picture, for this electron exchange to occur, the process requires an exchange time comparable to the inverse energy gap between the two lowest-energy states of $H_2^+$ along the characteristic target-projectile separation during the scattering process. In Fig.~\ref{fig:nucel}(f) we calculate the probability that the target electron unbounds from the parent nucleus. As we choose a nuclear tunneling closer to the electronic component ($J_n/J_e>0.1$), the kinetic energy of the projectile greatly exceeds the ionization energy of the target electron $(K_p/I\gg100)$, observing that the associated short interaction time suppresses further ionization events.

\begin{figure}[!tbp]
    \centering
    \includegraphics[width = 1\columnwidth]{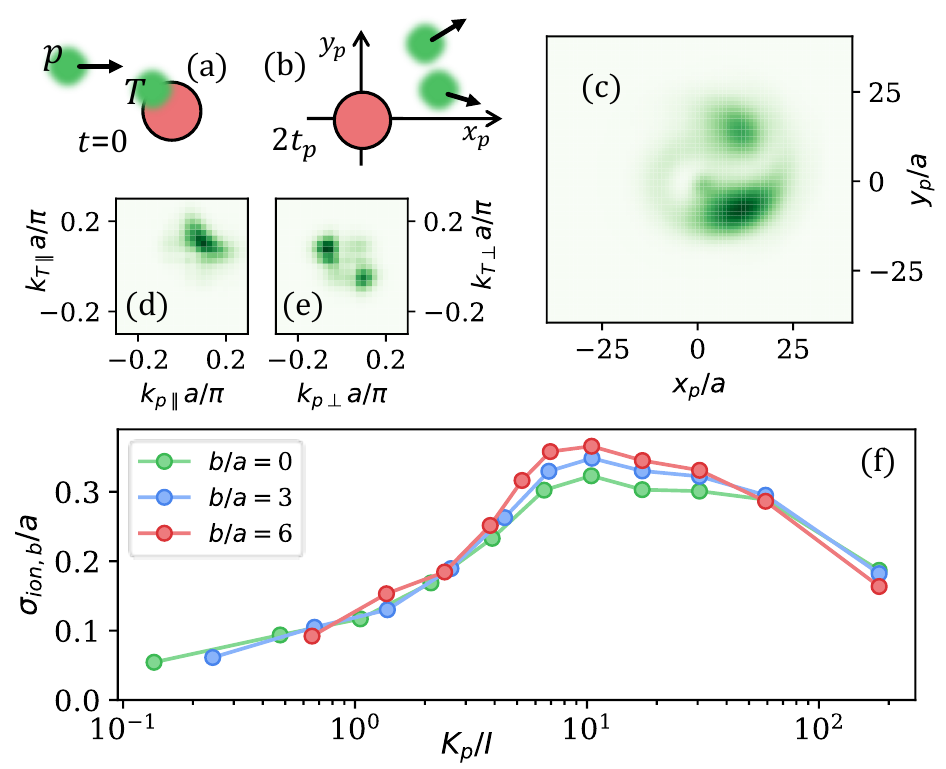}
\caption{(a,b) Scheme of the electronic scattering against simulated hydrogen at initial and final time $2t_p$, respectively. (c) Projectile probability density at time $2t_p$, conditioned to the ionization of the target. (d,e) Momentum correlations between the projectile [p] and target [T] electron at time $2t_p$ along the incident direction and the orthogonal axis, respectively, for $k_0a=0.17\pi$ and $b/a=6$. (f) Total ionization cross section for distinguishable electrons and a single scattering event with impact factor $b$. \emph{Parameters:} $V_0/J_e=1$, $w_0/a=4.25$, $r_0/a=1.5$, $N_{x,y}=80$.}
\label{fig:elel}
\end{figure}

\noindent \emph{Simulating inelastic ionization.-} Now, we investigate the case in which an electron is launched against a target hydrogen atom. For now, we consider that the electrons involved have opposite spin, so that they are distinguishable particles, and that the nuclear potential is fixed [see the scheme in Figs.~\ref{fig:elel}(a,b)]. Now that the target and projectile electrons have the same simulated mass, they present the same tunneling rate $J_e$, which forces us to control the incoming kinetic energy $K_p$ through the carrier wavevector $k_0$. 
In Fig.~\ref{fig:elel}(c) we show the scattered electron at final time $2t_p$. As confirmed in Figs.~\ref{fig:elel}(d,e), when the target electron is ejected by the incoming projectile, both electrons are mostly emitted in the forward direction, while an anticorrelated momentum in the orthogonal axis is caused by their electronic repulsion.

The ionization cross section $\sigma\stx{ion}$ accounts for the amount of scattering events in which the target electron is ejected from the atom (see Ref.~\cite{EM}). In Fig.~\ref{fig:elel}(f) we calculate the ionization cross section as a function of the impact parameter for different values of the incoming kinetic energy, defined by the momentum of the wavepacket. For incoming energies below the ionization threshold, $I$, no ionization should occur, and we observe finite-size effects due to the larger initial wavepackets used to access this region. In the limit $K_p \gg I$, the short interaction time greatly reduces the scattering events and the ionization is suppressed. The maximum ionization cross section thus corresponds to $K_p \sim I$, as observed in conventional scattering experiments~\cite{fiteCollisions1958,rudgeTheory1968,bransden2003physics}. 

\emph{Conclusions \& outlook.-} 
To sum up, we have shown that ultracold molecules moving in two-dimensional optical lattices can be used to simulate simplified chemistry models where both the electronic and nuclear degrees of freedom are preserved. We have observed that the natural cubic scaling of dipolar interactions enables one to access phenomena where the interactions among electrons and nuclei are relevant, as is the case of scattering events with electronic exchange or inelastic ionization. Compared with scattering experiments with real gases, the simulated dynamics occurs at a more favorable spatial and temporal scale that can be measured with atomic gas microscopy~\cite{Gross2021, Weitenberg2011}. We foresee that as the number of simulated particles increases, this unprecedented access to single-particle events can thus provide a complementary tool to understand and benchmark numerical methods in scattering regimes inaccessible by classical methods, as proposed in other fields such as lattice gauge theories~\cite{bennewitzSimulating2024, sharmaMeson2025, suColdAtom2024}. Other relevant examples are molecular configurations with exceptional points that require one to analyze the geometric phase of individual trajectories~\cite{domckeConical2004,meadDetermination1979,wuPrediction1993,althorpeObservation2002}.
Finally, while in this Letter electrons and nuclei are codified by different rotational levels of a molecule, one can also consider other alternatives, such as relying on different states of paramagnetic~\cite{norciaTwodimensional2021,baierExtended2016,chomazDipolar2022}
or Rydberg atoms~\cite{balewskiRydberg2014, guardado-sanchezQuench2021}.

\begin{acknowledgements}
 \emph{Acknowledgements.-} 
  We acknowledge useful discussions with Octavio Roncero and Arthur Christianen about the simulation of molecular dynamics.
J. A.-L. acknowledges support by the Spanish Ministerio de Ciencia, Innovación y Universidades (grant PID2023-147469NB-C21, financed by MICIU/AEI/10.13039/501100011033 and FEDER-EU).
 A.G.-T. acknowledges support from the CSIC Research Platform on Quantum Technologies PTI-001, from Spanish projects PID2021-127968NB-I00 funded by MICIU/AEI/10.13039/501100011033/ and by FEDER Una manera de hacer Europa, and from the QUANTERA project MOLAR with reference PCI2024-153449 and funded MICIU/AEI/10.13039/501100011033 and by the European Union. 
 J.I.C. acknowledges funding from the project FermiQP of the Bildungsministerium für Bildung und Forschung (BMBF) as well as from the Munich Quantum Valley, which is supported by the Bavarian State Government with funds from the High tech Agenda Bayern Plus.
\end{acknowledgements}

\clearpage
\newpage
\appendix
\renewcommand\thefigure{E\arabic{figure}}  
\renewcommand{\theequation}{E\arabic{equation}}
\renewcommand\appendixname{EM}
\setcounter{figure}{0} 
\setcounter{equation}{0} 
\section*{End Matter}
\section{Further details about the dipolar interaction}
In the effective interaction depicted in Eq.~\eqref{eq:Vmol}, one can observe that the resulting potential $V_{\sigma,\sigma'}(r)$ between two molecules in rotational states $\sigma$ and $\sigma'$ separated by distance $r$ reads as,
\begin{align}
    V_{\uparrow \uparrow}(r)&=(U_z/4+W)/r^3\\
    V_{\uparrow \downarrow}(r)&=-U_z/(4r^3)\\
    V_{\downarrow \downarrow}(r)&=(U_z/4-W)/r^3
\end{align}
for negligible mixing $U_\bot/U_z\ll 1$. The condition, $4\abs{W}<U_z$, thus ensures that molecules in the same state repel each other, while those in different states get attracted, as it occurs for $\beta\stx{DC}\sim 8$ with the parameters explored in Fig.~\ref{fig:scheme}. Therefore, $W$  does not change the scaling of the interactions, but only their strength, which can be used to further engineer the stronger attractive and repulsive forces associated with nuclei with different atomic numbers.

Regarding the undesired mixing of states $\ket{\uparrow}$ and $\ket{\downarrow}$ with other rotational levels that may originate from the dipolar interaction, this is conveniently suppressed by energy detunings dictated by the rotational constant. As $B_N$ is typically on the order of a few GHz, this energy gap is much larger than the dipolar interaction between molecules separated by typical optical lattice spacings $r\sim$ 500 nm, and also much larger than ultracold temperatures~\cite{wall2015Quantum}. 

\section{Wavepacket engineering}
In this Letter we have focused on the simulation of a scattering process where an incoming particle with initial momentum $k_0$ along the x-axis collides with a target species. First, let us consider that the projectile is a single particle (a nucleus or an electron), represented by the corresponding rotational level of a molecule. Initially, it is prepared in the spatially-gaussian ground state of width $w_0$ for a harmonic potential created, e.g. by an external optical potential. For a one-dimensional wavepacket to move toward its target, one can use a spatial-light modulator to imprint the needed site-dependent phase, which results in
\begin{equation}
    \psi_{w_0,k_0}(x)=\frac{1}{(\pi w_0^2)^{1/4}}e^{-\frac{(x-x_0)^2}{2w_0^2}}e^{ik_0 x}\,.
    \label{eq:psi0}
\end{equation}
One should observe that, for the nearest-neighbor tunneling in Eq.~\eqref{eq:Htarget}, the dispersion relation for this wavepacket is of the form $\omega(k)=-2J \cos(k a)$. The center of the wavepacket thus propagates as $x_0(t)=x_0+v_gt$, with group velocity $v_g(k_0)=\partial_k \omega(k_0)$, so that the fastest propagation corresponds to the carrier wavenumber $k_0 a=\pi/2$.

In addition to this propagation, the projectile can diffuse  due to the lowest-order quadratic expansion, $\Gamma(k_0)=\partial^2_k \omega(k_0)=2Ja^2\cos(k_0a) $. As a consequence, the width of the wavepacket, $w_{J,k_0}$ increases over time, and an additional spatial-dependent phase shift $\varphi_{J,k_0}$ appears~\cite{schwartz2016lecture}:
\begin{align}
    w_{J,k_0}(t)&= w_0\sqrt{1+\frac{[\Gamma(k_0) t]^2}{w_0^4}}\,,\label{eq:wdt} \\
    \varphi_{J,k_0}(x,t)&= \frac{[x-x_0(t)]^2}{2}\frac{\Gamma(k_0) t}{w_0^4+[\Gamma(k_0) t]^2 }\label{eq:Jdt}\,.
\end{align}
For 2D wavepackets with carrier wavevector $\kk_0=(k_0,0)$, the dominant distortion of the wavepacket in the direction orthogonal to the propagation may be undesirable, as one would like to preserve its symmetric shape when the collision occurs. 
To prevent this undesired expansion of the wavepacket, we initiate the simulation with self-focusing wavepackets that compensate for these additional effects. 
\begin{equation}
\label{eq:selfFocus}
    \psi\stx{in}(\rr,0)=\prod_{\xi\in \{x,y\}} \psi_{w_{J,k_{0\xi}}(t_p),k_{0\xi}}(r_\xi) e^{-i \varphi_{J,k_{0\xi}}(r_\xi,t_p)}\,.
\end{equation}
Regarding the next terms of the dispersion relation, they will introduce an unwanted skewness on the propagation that increases over time. For this effect to be controlled along the propagation time $t_p=s/v_g$ needed to reach the target at horizontal distance $s$, this translates into a minimal initial width for the wavepacket, $w_0/a\gtrsim (s/a)^{1/3}$. Under this condition, the state is initially localized in momentum, and the linear dispersion relation is a good approximation for the propagation of the wavepacket.

\section{Units mapping}
The ionization energy ($I$) and the average radius ($r_B$), define the natural energy and length scales of the simulated hydrogen atom, respectively. They are conveniently defined as the expected energy and radius of the electronic ground state: $I=-\ev{H}$ and $r_B=\sqrt{\ev{r^2}}$.

\begin{figure}[tb]
    \centering
    \includegraphics[width = \columnwidth]{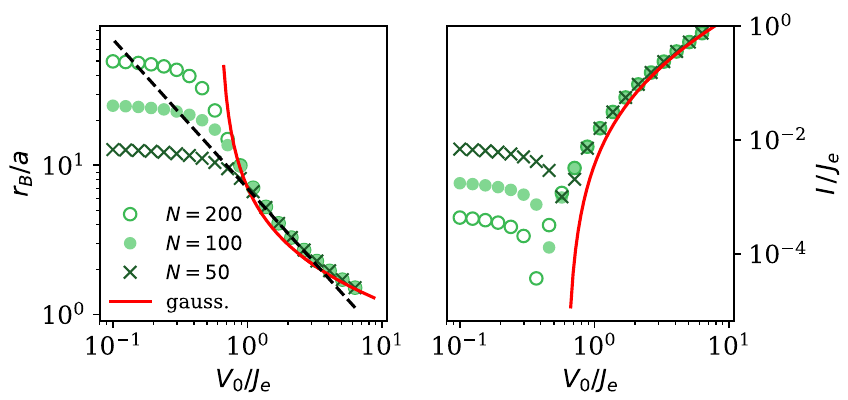}
\caption{(a) Expected radius $r_B$, and (b) ionization energy $I$, for a simulated hydrogen atom with nuclear potential $V\stx{mol}(V_0,r)$ with $r_0/a=1.5$, as a function of the potential strength $V_0$ and different values of the sites per side of the lattice $N$ (see legend). Dashed line follows the scaling $r_B/a \propto J_e/V_0$, and red line shows the expectation for a gaussian ansatz.}
\label{fig:atomic}
\end{figure}

In Fig.~\ref{fig:atomic} we calculate $I$ and $r_B$ for the attractive nuclear potential, $V\stx{mol}(r)$, as a function of the ratio $V_0/J_e$ for lattices with different numbers of sites. For $V_0\ll J_e$, the electronic state spreads across the entire lattice and becomes sensitive to finite-size effects, where smaller lattices are more affected. In the opposite limit, one is more affected by the discretization of the lattice space. To further characterize this limit, one can consider gaussian states of the form $\psi_g(\rr,w)=(\sqrt{\pi}w)^{-1}e^{-r^2/(2w^2)}$, to have an analytic approximation for the energy of the ground state as a function of the width $w$ of the state. To obtain the red lines in Fig.~\ref{fig:atomic}, we numerically minimize the energy for the gaussian ansatz and the potential $V\stx{mol}$, where the optimal width satisfies $w\stx{min}=r_B$. For the chosen cutoff length, $r_0/a=1.5$, we observe that the potential only allows one bound state, and that $r_B/a\sim 10$ is the maximum average radius one can simulate in a lattice with $N_{x,y}\sim 100$ sites per side before finite-size effects appear, which corresponds to the choice $V_0\sim J_e$ considered in Figs.~\ref{fig:nucel} and~\ref{fig:elel}.

\section{Numerical methods}
To numerically calculate the temporal evolution of the incoming self-focusing wavepacket~\eqref{eq:selfFocus} and its interaction with the target molecule, we use the split method~\cite{Lloyd1996,childsTheory2021}. In this approach, we take advantage of the fact that the first line in Hamiltonian~\eqref{eq:Htarget} ($\hat H_k)$ is diagonal in momentum space, while the second line is diagonal in real space ($\hat V)$. The evolution under the total Hamiltonian $\hat H=\hat H_k+\hat V$ is then Trotterized in short temporal intervals $\tau$ as,
$$
e^{-i\hat H t}=\prod_{i=1}^{n=t/\tau}\pa{e^{-i\hat V \tau} e^{-i\hat H_k \tau}} \,,
$$
whose associated error is of order $\mathcal{O}\pa{\tau^2n|[\hat V,\hat H_k]| }$. Before applying each evolution operator, one can then perform a fast Fourier transformation of the evolving state to conveniently express it in the appropriate real or momentum basis. This strategy highly reduces the computational cost of the operation, as both the state and operators have the size of the Hilbert space (as compared with the quadratic size that operators would require in an inconvenient basis). In this work, we have used $J_e \tau=0.5$, where convergence is observed until the final time $2t_p$. The ionization rate in Fig.~\ref{fig:nucel}(f) corresponds to the probability that this final state is orthogonal to both the bound target and the unscattered projectile.

To extract numerically the differential cross section once the final time $2t_p$ is reached, one can subtract the contribution of the freely propagating wavepacket from the evolved state to calculate the scattering wavefunction $\psi\stx{sc,b}=\psi(\rr)-\psi\stx{in}(\rr)$. Then, the overlap of the resulting density probability $\abs{\psi\stx{sc,b}}^2$ with angular probe functions $P_b(\theta_0) =  \bra{\psi\stx{sc,b}} e^{-(\theta-\theta_0)^2/\Delta\theta^2} \ket{\psi\stx{sc,b}}$ provides the probability of the projectile being scattered at an angle $\theta$ after hitting the target, which allows one to reconstruct the differential scattering cross-section. 

For our single-trajectory simulation, the total inelastic cross section $\sigma\stx{ion,b}$ can be calculated as the ratio between the number of ionization  events $|\psi\stx{ion,b}|^2$ and the number of targets per unit length, which we estimate as $n_T=w_0^{-1}\int d\rr |\psi_g(\rr,r_B)|^2 e^{-(y-b)^2/2w_0^2}$. Therefore, the inelastic scattering cross section represented in Fig.~\ref{fig:elel}(f) is evaluated as $\sigma\stx{ion,b}=\tilde r_B \exp\co{b^2/\tilde r_B^2} |\psi\stx{ion,b}|^2$, where $\tilde r_B=\sqrt{r_B^2+2w_0^2}$. For the calculation of the ionization state, $\psi\stx{ion,b}$, we ensure that inelastic ionization has occurred using the contribution of the scattered state where both electrons are orthogonal to the bound state of the fixed nuclear potential.

\begin{thebibliography}{76}%
\makeatletter
\providecommand \@ifxundefined [1]{%
 \@ifx{#1\undefined}
}%
\providecommand \@ifnum [1]{%
 \ifnum #1\expandafter \@firstoftwo
 \else \expandafter \@secondoftwo
 \fi
}%
\providecommand \@ifx [1]{%
 \ifx #1\expandafter \@firstoftwo
 \else \expandafter \@secondoftwo
 \fi
}%
\providecommand \natexlab [1]{#1}%
\providecommand \enquote  [1]{``#1''}%
\providecommand \bibnamefont  [1]{#1}%
\providecommand \bibfnamefont [1]{#1}%
\providecommand \citenamefont [1]{#1}%
\providecommand \href@noop [0]{\@secondoftwo}%
\providecommand \href [0]{\begingroup \@sanitize@url \@href}%
\providecommand \@href[1]{\@@startlink{#1}\@@href}%
\providecommand \@@href[1]{\endgroup#1\@@endlink}%
\providecommand \@sanitize@url [0]{\catcode `\\12\catcode `\$12\catcode `\&12\catcode `\#12\catcode `\^12\catcode `\_12\catcode `\%12\relax}%
\providecommand \@@startlink[1]{}%
\providecommand \@@endlink[0]{}%
\providecommand \url  [0]{\begingroup\@sanitize@url \@url }%
\providecommand \@url [1]{\endgroup\@href {#1}{\urlprefix }}%
\providecommand \urlprefix  [0]{URL }%
\providecommand \Eprint [0]{\href }%
\providecommand \doibase [0]{https://doi.org/}%
\providecommand \selectlanguage [0]{\@gobble}%
\providecommand \bibinfo  [0]{\@secondoftwo}%
\providecommand \bibfield  [0]{\@secondoftwo}%
\providecommand \translation [1]{[#1]}%
\providecommand \BibitemOpen [0]{}%
\providecommand \bibitemStop [0]{}%
\providecommand \bibitemNoStop [0]{.\EOS\space}%
\providecommand \EOS [0]{\spacefactor3000\relax}%
\providecommand \BibitemShut  [1]{\csname bibitem#1\endcsname}%
\let\auto@bib@innerbib\@empty
\bibitem [{\citenamefont {Lewenstein}\ \emph {et~al.}(2007)\citenamefont {Lewenstein}, \citenamefont {Sanpera}, \citenamefont {Ahufinger}, \citenamefont {Damski}, \citenamefont {Sen(De)},\ and\ \citenamefont {Sen}}]{Lewenstein2007}%
  \BibitemOpen
  \bibfield  {author} {\bibinfo {author} {\bibfnamefont {M.}~\bibnamefont {Lewenstein}}, \bibinfo {author} {\bibfnamefont {A.}~\bibnamefont {Sanpera}}, \bibinfo {author} {\bibfnamefont {V.}~\bibnamefont {Ahufinger}}, \bibinfo {author} {\bibfnamefont {B.}~\bibnamefont {Damski}}, \bibinfo {author} {\bibfnamefont {A.}~\bibnamefont {Sen(De)}},\ and\ \bibinfo {author} {\bibfnamefont {U.}~\bibnamefont {Sen}},\ }\bibfield  {title} {\bibinfo {title} {Ultracold atomic gases in optical lattices: Mimicking condensed matter physics and beyond},\ }\href {https://doi.org/10.1080/00018730701223200} {\bibfield  {journal} {\bibinfo  {journal} {Advances in Physics}\ }\textbf {\bibinfo {volume} {56}},\ \bibinfo {pages} {243} (\bibinfo {year} {2007})}\BibitemShut {NoStop}%
\bibitem [{\citenamefont {Gross}\ and\ \citenamefont {Bloch}(2017)}]{Gross2017}%
  \BibitemOpen
  \bibfield  {author} {\bibinfo {author} {\bibfnamefont {C.}~\bibnamefont {Gross}}\ and\ \bibinfo {author} {\bibfnamefont {I.}~\bibnamefont {Bloch}},\ }\bibfield  {title} {\bibinfo {title} {Quantum simulations with ultracold atoms in optical lattices.},\ }\href {https://doi.org/10.1126/science.aal3837} {\bibfield  {journal} {\bibinfo  {journal} {Science (New York, N.Y.)}\ }\textbf {\bibinfo {volume} {357}},\ \bibinfo {pages} {995} (\bibinfo {year} {2017})},\ \Eprint {https://arxiv.org/abs/28883070} {28883070} \BibitemShut {NoStop}%
\bibitem [{\citenamefont {Bloch}\ \emph {et~al.}(2012)\citenamefont {Bloch}, \citenamefont {Dalibard},\ and\ \citenamefont {Nascimb{\`e}ne}}]{Bloch2012}%
  \BibitemOpen
  \bibfield  {author} {\bibinfo {author} {\bibfnamefont {I.}~\bibnamefont {Bloch}}, \bibinfo {author} {\bibfnamefont {J.}~\bibnamefont {Dalibard}},\ and\ \bibinfo {author} {\bibfnamefont {S.}~\bibnamefont {Nascimb{\`e}ne}},\ }\bibfield  {title} {\bibinfo {title} {Quantum simulations with ultracold quantum gases},\ }\href {https://doi.org/10.1038/nphys2259} {\bibfield  {journal} {\bibinfo  {journal} {Nature Physics}\ }\textbf {\bibinfo {volume} {8}},\ \bibinfo {pages} {267} (\bibinfo {year} {2012})}\BibitemShut {NoStop}%
\bibitem [{\citenamefont {Gross}\ and\ \citenamefont {Bakr}(2021{\natexlab{a}})}]{Gross_2021}%
  \BibitemOpen
  \bibfield  {author} {\bibinfo {author} {\bibfnamefont {C.}~\bibnamefont {Gross}}\ and\ \bibinfo {author} {\bibfnamefont {W.~S.}\ \bibnamefont {Bakr}},\ }\bibfield  {title} {\bibinfo {title} {Quantum gas microscopy for single atom and spin detection},\ }\href {https://doi.org/10.1038/s41567-021-01370-5} {\bibfield  {journal} {\bibinfo  {journal} {Nature Physics}\ }\textbf {\bibinfo {volume} {17}},\ \bibinfo {pages} {1316} (\bibinfo {year} {2021}{\natexlab{a}})}\BibitemShut {NoStop}%
\bibitem [{\citenamefont {Hart}\ \emph {et~al.}(2015)\citenamefont {Hart}, \citenamefont {Duarte}, \citenamefont {Yang}, \citenamefont {Liu}, \citenamefont {Paiva}, \citenamefont {Khatami}, \citenamefont {Scalettar}, \citenamefont {Trivedi}, \citenamefont {Huse},\ and\ \citenamefont {Hulet}}]{Hart_2015}%
  \BibitemOpen
  \bibfield  {author} {\bibinfo {author} {\bibfnamefont {R.~A.}\ \bibnamefont {Hart}}, \bibinfo {author} {\bibfnamefont {P.~M.}\ \bibnamefont {Duarte}}, \bibinfo {author} {\bibfnamefont {T.-L.}\ \bibnamefont {Yang}}, \bibinfo {author} {\bibfnamefont {X.}~\bibnamefont {Liu}}, \bibinfo {author} {\bibfnamefont {T.}~\bibnamefont {Paiva}}, \bibinfo {author} {\bibfnamefont {E.}~\bibnamefont {Khatami}}, \bibinfo {author} {\bibfnamefont {R.~T.}\ \bibnamefont {Scalettar}}, \bibinfo {author} {\bibfnamefont {N.}~\bibnamefont {Trivedi}}, \bibinfo {author} {\bibfnamefont {D.~A.}\ \bibnamefont {Huse}},\ and\ \bibinfo {author} {\bibfnamefont {R.~G.}\ \bibnamefont {Hulet}},\ }\bibfield  {title} {\bibinfo {title} {Observation of antiferromagnetic correlations in the hubbard model with ultracold atoms},\ }\href {https://doi.org/10.1038/nature14223} {\bibfield  {journal} {\bibinfo  {journal} {Nature}\ }\textbf {\bibinfo {volume} {519}},\ \bibinfo {pages} {211} (\bibinfo {year} {2015})}\BibitemShut {NoStop}%
\bibitem [{\citenamefont {Boll}\ \emph {et~al.}(2016)\citenamefont {Boll}, \citenamefont {Hilker}, \citenamefont {Salomon}, \citenamefont {Omran}, \citenamefont {Nespolo}, \citenamefont {Pollet}, \citenamefont {Bloch},\ and\ \citenamefont {Gross}}]{Boll_2016}%
  \BibitemOpen
  \bibfield  {author} {\bibinfo {author} {\bibfnamefont {M.}~\bibnamefont {Boll}}, \bibinfo {author} {\bibfnamefont {T.~A.}\ \bibnamefont {Hilker}}, \bibinfo {author} {\bibfnamefont {G.}~\bibnamefont {Salomon}}, \bibinfo {author} {\bibfnamefont {A.}~\bibnamefont {Omran}}, \bibinfo {author} {\bibfnamefont {J.}~\bibnamefont {Nespolo}}, \bibinfo {author} {\bibfnamefont {L.}~\bibnamefont {Pollet}}, \bibinfo {author} {\bibfnamefont {I.}~\bibnamefont {Bloch}},\ and\ \bibinfo {author} {\bibfnamefont {C.}~\bibnamefont {Gross}},\ }\bibfield  {title} {\bibinfo {title} {Spin- and density-resolved microscopy of antiferromagnetic correlations in fermi-hubbard chains},\ }\href {https://doi.org/10.1126/science.aag1635} {\bibfield  {journal} {\bibinfo  {journal} {Science}\ }\textbf {\bibinfo {volume} {353}},\ \bibinfo {pages} {1257} (\bibinfo {year} {2016})}\BibitemShut {NoStop}%
\bibitem [{\citenamefont {Cheuk}\ \emph {et~al.}(2016)\citenamefont {Cheuk}, \citenamefont {Nichols}, \citenamefont {Lawrence}, \citenamefont {Okan}, \citenamefont {Zhang}, \citenamefont {Khatami}, \citenamefont {Trivedi}, \citenamefont {Paiva}, \citenamefont {Rigol},\ and\ \citenamefont {Zwierlein}}]{Cheuk_2016}%
  \BibitemOpen
  \bibfield  {author} {\bibinfo {author} {\bibfnamefont {L.~W.}\ \bibnamefont {Cheuk}}, \bibinfo {author} {\bibfnamefont {M.~A.}\ \bibnamefont {Nichols}}, \bibinfo {author} {\bibfnamefont {K.~R.}\ \bibnamefont {Lawrence}}, \bibinfo {author} {\bibfnamefont {M.}~\bibnamefont {Okan}}, \bibinfo {author} {\bibfnamefont {H.}~\bibnamefont {Zhang}}, \bibinfo {author} {\bibfnamefont {E.}~\bibnamefont {Khatami}}, \bibinfo {author} {\bibfnamefont {N.}~\bibnamefont {Trivedi}}, \bibinfo {author} {\bibfnamefont {T.}~\bibnamefont {Paiva}}, \bibinfo {author} {\bibfnamefont {M.}~\bibnamefont {Rigol}},\ and\ \bibinfo {author} {\bibfnamefont {M.~W.}\ \bibnamefont {Zwierlein}},\ }\bibfield  {title} {\bibinfo {title} {Observation of spatial charge and spin correlations in the 2d fermi-hubbard model},\ }\href {https://doi.org/10.1126/science.aag3349} {\bibfield  {journal} {\bibinfo  {journal} {Science}\ }\textbf {\bibinfo {volume} {353}},\ \bibinfo {pages} {1260} (\bibinfo {year} {2016})}\BibitemShut {NoStop}%
\bibitem [{\citenamefont {Mazurenko}\ \emph {et~al.}(2017)\citenamefont {Mazurenko}, \citenamefont {Chiu}, \citenamefont {Ji}, \citenamefont {Parsons}, \citenamefont {Kan{\'a}sz-Nagy}, \citenamefont {Schmidt}, \citenamefont {Grusdt}, \citenamefont {Demler}, \citenamefont {Greif},\ and\ \citenamefont {Greiner}}]{Mazurenko_2017}%
  \BibitemOpen
  \bibfield  {author} {\bibinfo {author} {\bibfnamefont {A.}~\bibnamefont {Mazurenko}}, \bibinfo {author} {\bibfnamefont {C.~S.}\ \bibnamefont {Chiu}}, \bibinfo {author} {\bibfnamefont {G.}~\bibnamefont {Ji}}, \bibinfo {author} {\bibfnamefont {M.~F.}\ \bibnamefont {Parsons}}, \bibinfo {author} {\bibfnamefont {M.}~\bibnamefont {Kan{\'a}sz-Nagy}}, \bibinfo {author} {\bibfnamefont {R.}~\bibnamefont {Schmidt}}, \bibinfo {author} {\bibfnamefont {F.}~\bibnamefont {Grusdt}}, \bibinfo {author} {\bibfnamefont {E.}~\bibnamefont {Demler}}, \bibinfo {author} {\bibfnamefont {D.}~\bibnamefont {Greif}},\ and\ \bibinfo {author} {\bibfnamefont {M.}~\bibnamefont {Greiner}},\ }\bibfield  {title} {\bibinfo {title} {A cold-atom fermi--hubbard antiferromagnet},\ }\href {https://doi.org/10.1038/nature22362} {\bibfield  {journal} {\bibinfo  {journal} {Nature}\ }\textbf {\bibinfo {volume} {545}},\ \bibinfo {pages} {462} (\bibinfo {year} {2017})}\BibitemShut {NoStop}%
\bibitem [{\citenamefont {Xu}\ \emph {et~al.}(2023)\citenamefont {Xu}, \citenamefont {Kendrick}, \citenamefont {Kale}, \citenamefont {Gang}, \citenamefont {Ji}, \citenamefont {Scalettar}, \citenamefont {Lebrat},\ and\ \citenamefont {Greiner}}]{Xu_2023}%
  \BibitemOpen
  \bibfield  {author} {\bibinfo {author} {\bibfnamefont {M.}~\bibnamefont {Xu}}, \bibinfo {author} {\bibfnamefont {L.~H.}\ \bibnamefont {Kendrick}}, \bibinfo {author} {\bibfnamefont {A.}~\bibnamefont {Kale}}, \bibinfo {author} {\bibfnamefont {Y.}~\bibnamefont {Gang}}, \bibinfo {author} {\bibfnamefont {G.}~\bibnamefont {Ji}}, \bibinfo {author} {\bibfnamefont {R.~T.}\ \bibnamefont {Scalettar}}, \bibinfo {author} {\bibfnamefont {M.}~\bibnamefont {Lebrat}},\ and\ \bibinfo {author} {\bibfnamefont {M.}~\bibnamefont {Greiner}},\ }\bibfield  {title} {\bibinfo {title} {Frustration- and doping-induced magnetism in a fermi--hubbard simulator},\ }\href {https://doi.org/10.1038/s41586-023-06280-5} {\bibfield  {journal} {\bibinfo  {journal} {Nature}\ }\textbf {\bibinfo {volume} {620}},\ \bibinfo {pages} {971} (\bibinfo {year} {2023})}\BibitemShut {NoStop}%
\bibitem [{\citenamefont {Shao}\ \emph {et~al.}(2024)\citenamefont {Shao}, \citenamefont {Wang}, \citenamefont {Zhu}, \citenamefont {Zhu}, \citenamefont {Sun}, \citenamefont {Chen}, \citenamefont {Zhang}, \citenamefont {Fan}, \citenamefont {Deng}, \citenamefont {Yao}, \citenamefont {Chen},\ and\ \citenamefont {Pan}}]{Shao_2024}%
  \BibitemOpen
  \bibfield  {author} {\bibinfo {author} {\bibfnamefont {H.-J.}\ \bibnamefont {Shao}}, \bibinfo {author} {\bibfnamefont {Y.-X.}\ \bibnamefont {Wang}}, \bibinfo {author} {\bibfnamefont {D.-Z.}\ \bibnamefont {Zhu}}, \bibinfo {author} {\bibfnamefont {Y.-S.}\ \bibnamefont {Zhu}}, \bibinfo {author} {\bibfnamefont {H.-N.}\ \bibnamefont {Sun}}, \bibinfo {author} {\bibfnamefont {S.-Y.}\ \bibnamefont {Chen}}, \bibinfo {author} {\bibfnamefont {C.}~\bibnamefont {Zhang}}, \bibinfo {author} {\bibfnamefont {Z.-J.}\ \bibnamefont {Fan}}, \bibinfo {author} {\bibfnamefont {Y.}~\bibnamefont {Deng}}, \bibinfo {author} {\bibfnamefont {X.-C.}\ \bibnamefont {Yao}}, \bibinfo {author} {\bibfnamefont {Y.-A.}\ \bibnamefont {Chen}},\ and\ \bibinfo {author} {\bibfnamefont {J.-W.}\ \bibnamefont {Pan}},\ }\bibfield  {title} {\bibinfo {title} {Antiferromagnetic phase transition in a 3d fermionic hubbard model},\ }\href {https://doi.org/10.1038/s41586-024-07689-2} {\bibfield  {journal} {\bibinfo  {journal} {Nature}\ }\textbf {\bibinfo
  {volume} {632}},\ \bibinfo {pages} {267} (\bibinfo {year} {2024})}\BibitemShut {NoStop}%
\bibitem [{\citenamefont {Chiu}\ \emph {et~al.}(2019)\citenamefont {Chiu}, \citenamefont {Ji}, \citenamefont {Bohrdt}, \citenamefont {Xu}, \citenamefont {Knap}, \citenamefont {Demler}, \citenamefont {Grusdt}, \citenamefont {Greiner},\ and\ \citenamefont {Greif}}]{Chiu_2019}%
  \BibitemOpen
  \bibfield  {author} {\bibinfo {author} {\bibfnamefont {C.~S.}\ \bibnamefont {Chiu}}, \bibinfo {author} {\bibfnamefont {G.}~\bibnamefont {Ji}}, \bibinfo {author} {\bibfnamefont {A.}~\bibnamefont {Bohrdt}}, \bibinfo {author} {\bibfnamefont {M.}~\bibnamefont {Xu}}, \bibinfo {author} {\bibfnamefont {M.}~\bibnamefont {Knap}}, \bibinfo {author} {\bibfnamefont {E.}~\bibnamefont {Demler}}, \bibinfo {author} {\bibfnamefont {F.}~\bibnamefont {Grusdt}}, \bibinfo {author} {\bibfnamefont {M.}~\bibnamefont {Greiner}},\ and\ \bibinfo {author} {\bibfnamefont {D.}~\bibnamefont {Greif}},\ }\bibfield  {title} {\bibinfo {title} {String patterns in the doped hubbard model},\ }\href {https://doi.org/10.1126/science.aav3587} {\bibfield  {journal} {\bibinfo  {journal} {Science}\ }\textbf {\bibinfo {volume} {365}},\ \bibinfo {pages} {251} (\bibinfo {year} {2019})}\BibitemShut {NoStop}%
\bibitem [{\citenamefont {Bourgund}\ \emph {et~al.}(2025)\citenamefont {Bourgund}, \citenamefont {Chalopin}, \citenamefont {Bojovi{\'c}}, \citenamefont {Schl{\"o}mer}, \citenamefont {Wang}, \citenamefont {Franz}, \citenamefont {Hirthe}, \citenamefont {Bohrdt}, \citenamefont {Grusdt}, \citenamefont {Bloch} \emph {et~al.}}]{Bourgund_2023}%
  \BibitemOpen
  \bibfield  {author} {\bibinfo {author} {\bibfnamefont {D.}~\bibnamefont {Bourgund}}, \bibinfo {author} {\bibfnamefont {T.}~\bibnamefont {Chalopin}}, \bibinfo {author} {\bibfnamefont {P.}~\bibnamefont {Bojovi{\'c}}}, \bibinfo {author} {\bibfnamefont {H.}~\bibnamefont {Schl{\"o}mer}}, \bibinfo {author} {\bibfnamefont {S.}~\bibnamefont {Wang}}, \bibinfo {author} {\bibfnamefont {T.}~\bibnamefont {Franz}}, \bibinfo {author} {\bibfnamefont {S.}~\bibnamefont {Hirthe}}, \bibinfo {author} {\bibfnamefont {A.}~\bibnamefont {Bohrdt}}, \bibinfo {author} {\bibfnamefont {F.}~\bibnamefont {Grusdt}}, \bibinfo {author} {\bibfnamefont {I.}~\bibnamefont {Bloch}}, \emph {et~al.},\ }\bibfield  {title} {\bibinfo {title} {Formation of individual stripes in a mixed-dimensional cold-atom fermi--hubbard system},\ }\href {https://doi.org/10.1038/s41586-024-08270-7}{\bibfield  {journal} {\bibinfo  {journal} {Nature}\ }\textbf {\bibinfo {volume} {637}},\ \bibinfo {pages} {57} (\bibinfo {year} {2025})}\BibitemShut {NoStop}%
\bibitem [{\citenamefont {Sompet}\ \emph {et~al.}(2022)\citenamefont {Sompet}, \citenamefont {Hirthe}, \citenamefont {Bourgund}, \citenamefont {Chalopin}, \citenamefont {Bibo}, \citenamefont {Koepsell}, \citenamefont {Bojovi{\'c}}, \citenamefont {Verresen}, \citenamefont {Pollmann}, \citenamefont {Salomon}, \citenamefont {Gross}, \citenamefont {Hilker},\ and\ \citenamefont {Bloch}}]{Sompet_2022}%
  \BibitemOpen
  \bibfield  {author} {\bibinfo {author} {\bibfnamefont {P.}~\bibnamefont {Sompet}}, \bibinfo {author} {\bibfnamefont {S.}~\bibnamefont {Hirthe}}, \bibinfo {author} {\bibfnamefont {D.}~\bibnamefont {Bourgund}}, \bibinfo {author} {\bibfnamefont {T.}~\bibnamefont {Chalopin}}, \bibinfo {author} {\bibfnamefont {J.}~\bibnamefont {Bibo}}, \bibinfo {author} {\bibfnamefont {J.}~\bibnamefont {Koepsell}}, \bibinfo {author} {\bibfnamefont {P.}~\bibnamefont {Bojovi{\'c}}}, \bibinfo {author} {\bibfnamefont {R.}~\bibnamefont {Verresen}}, \bibinfo {author} {\bibfnamefont {F.}~\bibnamefont {Pollmann}}, \bibinfo {author} {\bibfnamefont {G.}~\bibnamefont {Salomon}}, \bibinfo {author} {\bibfnamefont {C.}~\bibnamefont {Gross}}, \bibinfo {author} {\bibfnamefont {T.~A.}\ \bibnamefont {Hilker}},\ and\ \bibinfo {author} {\bibfnamefont {I.}~\bibnamefont {Bloch}},\ }\bibfield  {title} {\bibinfo {title} {Realizing the symmetry-protected haldane phase in fermi--hubbard ladders},\ }\href {https://doi.org/10.1038/s41586-022-04688-z}
  {\bibfield  {journal} {\bibinfo  {journal} {Nature}\ }\textbf {\bibinfo {volume} {606}},\ \bibinfo {pages} {484} (\bibinfo {year} {2022})}\BibitemShut {NoStop}%
\bibitem [{\citenamefont {Hirthe}\ \emph {et~al.}(2023)\citenamefont {Hirthe}, \citenamefont {Chalopin}, \citenamefont {Bourgund}, \citenamefont {Bojovi{\'c}}, \citenamefont {Bohrdt}, \citenamefont {Demler}, \citenamefont {Grusdt}, \citenamefont {Bloch},\ and\ \citenamefont {Hilker}}]{Hirthe_2023}%
  \BibitemOpen
  \bibfield  {author} {\bibinfo {author} {\bibfnamefont {S.}~\bibnamefont {Hirthe}}, \bibinfo {author} {\bibfnamefont {T.}~\bibnamefont {Chalopin}}, \bibinfo {author} {\bibfnamefont {D.}~\bibnamefont {Bourgund}}, \bibinfo {author} {\bibfnamefont {P.}~\bibnamefont {Bojovi{\'c}}}, \bibinfo {author} {\bibfnamefont {A.}~\bibnamefont {Bohrdt}}, \bibinfo {author} {\bibfnamefont {E.}~\bibnamefont {Demler}}, \bibinfo {author} {\bibfnamefont {F.}~\bibnamefont {Grusdt}}, \bibinfo {author} {\bibfnamefont {I.}~\bibnamefont {Bloch}},\ and\ \bibinfo {author} {\bibfnamefont {T.~A.}\ \bibnamefont {Hilker}},\ }\bibfield  {title} {\bibinfo {title} {Magnetically mediated hole pairing in fermionic ladders of ultracold atoms},\ }\href {https://doi.org/10.1038/s41586-022-05437-y} {\bibfield  {journal} {\bibinfo  {journal} {Nature}\ }\textbf {\bibinfo {volume} {613}},\ \bibinfo {pages} {463} (\bibinfo {year} {2023})}\BibitemShut {NoStop}%
\bibitem [{\citenamefont {Hartke}\ \emph {et~al.}(2023)\citenamefont {Hartke}, \citenamefont {Oreg}, \citenamefont {Turnbaugh}, \citenamefont {Jia},\ and\ \citenamefont {Zwierlein}}]{Hartke_2023}%
  \BibitemOpen
  \bibfield  {author} {\bibinfo {author} {\bibfnamefont {T.}~\bibnamefont {Hartke}}, \bibinfo {author} {\bibfnamefont {B.}~\bibnamefont {Oreg}}, \bibinfo {author} {\bibfnamefont {C.}~\bibnamefont {Turnbaugh}}, \bibinfo {author} {\bibfnamefont {N.}~\bibnamefont {Jia}},\ and\ \bibinfo {author} {\bibfnamefont {M.}~\bibnamefont {Zwierlein}},\ }\bibfield  {title} {\bibinfo {title} {Direct observation of nonlocal fermion pairing in an attractive fermi-hubbard gas},\ }\href {https://doi.org/10.1126/science.ade4245} {\bibfield  {journal} {\bibinfo  {journal} {Science}\ }\textbf {\bibinfo {volume} {381}},\ \bibinfo {pages} {82} (\bibinfo {year} {2023})}\BibitemShut {NoStop}%
\bibitem [{\citenamefont {Schreiber}\ \emph {et~al.}(2015)\citenamefont {Schreiber}, \citenamefont {Hodgman}, \citenamefont {Bordia}, \citenamefont {Lüschen}, \citenamefont {Fischer}, \citenamefont {Vosk}, \citenamefont {Altman}, \citenamefont {Schneider},\ and\ \citenamefont {Bloch}}]{Schreiber_2015}%
  \BibitemOpen
  \bibfield  {author} {\bibinfo {author} {\bibfnamefont {M.}~\bibnamefont {Schreiber}}, \bibinfo {author} {\bibfnamefont {S.~S.}\ \bibnamefont {Hodgman}}, \bibinfo {author} {\bibfnamefont {P.}~\bibnamefont {Bordia}}, \bibinfo {author} {\bibfnamefont {H.~P.}\ \bibnamefont {Lüschen}}, \bibinfo {author} {\bibfnamefont {M.~H.}\ \bibnamefont {Fischer}}, \bibinfo {author} {\bibfnamefont {R.}~\bibnamefont {Vosk}}, \bibinfo {author} {\bibfnamefont {E.}~\bibnamefont {Altman}}, \bibinfo {author} {\bibfnamefont {U.}~\bibnamefont {Schneider}},\ and\ \bibinfo {author} {\bibfnamefont {I.}~\bibnamefont {Bloch}},\ }\bibfield  {title} {\bibinfo {title} {Observation of many-body localization of interacting fermions in a quasirandom optical lattice},\ }\href {https://doi.org/10.1126/science.aaa7432} {\bibfield  {journal} {\bibinfo  {journal} {Science}\ }\textbf {\bibinfo {volume} {349}},\ \bibinfo {pages} {842} (\bibinfo {year} {2015})}\BibitemShut {NoStop}%
\bibitem [{\citenamefont {Nichols}\ \emph {et~al.}(2019)\citenamefont {Nichols}, \citenamefont {Cheuk}, \citenamefont {Okan}, \citenamefont {Hartke}, \citenamefont {Mendez}, \citenamefont {Senthil}, \citenamefont {Khatami}, \citenamefont {Zhang},\ and\ \citenamefont {Zwierlein}}]{Nichols_2019}%
  \BibitemOpen
  \bibfield  {author} {\bibinfo {author} {\bibfnamefont {M.~A.}\ \bibnamefont {Nichols}}, \bibinfo {author} {\bibfnamefont {L.~W.}\ \bibnamefont {Cheuk}}, \bibinfo {author} {\bibfnamefont {M.}~\bibnamefont {Okan}}, \bibinfo {author} {\bibfnamefont {T.~R.}\ \bibnamefont {Hartke}}, \bibinfo {author} {\bibfnamefont {E.}~\bibnamefont {Mendez}}, \bibinfo {author} {\bibfnamefont {T.}~\bibnamefont {Senthil}}, \bibinfo {author} {\bibfnamefont {E.}~\bibnamefont {Khatami}}, \bibinfo {author} {\bibfnamefont {H.}~\bibnamefont {Zhang}},\ and\ \bibinfo {author} {\bibfnamefont {M.~W.}\ \bibnamefont {Zwierlein}},\ }\bibfield  {title} {\bibinfo {title} {Spin transport in a mott insulator of ultracold fermions},\ }\href {https://doi.org/10.1126/science.aat4387} {\bibfield  {journal} {\bibinfo  {journal} {Science}\ }\textbf {\bibinfo {volume} {363}},\ \bibinfo {pages} {383} (\bibinfo {year} {2019})}\BibitemShut {NoStop}%
\bibitem [{\citenamefont {Guardado-Sanchez}\ \emph {et~al.}(2020)\citenamefont {Guardado-Sanchez}, \citenamefont {Morningstar}, \citenamefont {Spar}, \citenamefont {Brown}, \citenamefont {Huse},\ and\ \citenamefont {Bakr}}]{Guardado-Sanchez_2020}%
  \BibitemOpen
  \bibfield  {author} {\bibinfo {author} {\bibfnamefont {E.}~\bibnamefont {Guardado-Sanchez}}, \bibinfo {author} {\bibfnamefont {A.}~\bibnamefont {Morningstar}}, \bibinfo {author} {\bibfnamefont {B.~M.}\ \bibnamefont {Spar}}, \bibinfo {author} {\bibfnamefont {P.~T.}\ \bibnamefont {Brown}}, \bibinfo {author} {\bibfnamefont {D.~A.}\ \bibnamefont {Huse}},\ and\ \bibinfo {author} {\bibfnamefont {W.~S.}\ \bibnamefont {Bakr}},\ }\bibfield  {title} {\bibinfo {title} {Subdiffusion and heat transport in a tilted two-dimensional fermi-hubbard system},\ }\href {https://doi.org/10.1103/PhysRevX.10.011042} {\bibfield  {journal} {\bibinfo  {journal} {Phys. Rev. X}\ }\textbf {\bibinfo {volume} {10}},\ \bibinfo {pages} {011042} (\bibinfo {year} {2020})}\BibitemShut {NoStop}%
\bibitem [{\citenamefont {Scherg}\ \emph {et~al.}(2021)\citenamefont {Scherg}, \citenamefont {Kohlert}, \citenamefont {Sala}, \citenamefont {Pollmann}, \citenamefont {Hebbe~Madhusudhana}, \citenamefont {Bloch},\ and\ \citenamefont {Aidelsburger}}]{Scherg_2021}%
  \BibitemOpen
  \bibfield  {author} {\bibinfo {author} {\bibfnamefont {S.}~\bibnamefont {Scherg}}, \bibinfo {author} {\bibfnamefont {T.}~\bibnamefont {Kohlert}}, \bibinfo {author} {\bibfnamefont {P.}~\bibnamefont {Sala}}, \bibinfo {author} {\bibfnamefont {F.}~\bibnamefont {Pollmann}}, \bibinfo {author} {\bibfnamefont {B.}~\bibnamefont {Hebbe~Madhusudhana}}, \bibinfo {author} {\bibfnamefont {I.}~\bibnamefont {Bloch}},\ and\ \bibinfo {author} {\bibfnamefont {M.}~\bibnamefont {Aidelsburger}},\ }\bibfield  {title} {\bibinfo {title} {Observing non-ergodicity due to kinetic constraints in tilted fermi-hubbard chains},\ }\href {https://doi.org/10.1038/s41467-021-24726-0} {\bibfield  {journal} {\bibinfo  {journal} {Nature Communications}\ }\textbf {\bibinfo {volume} {12}},\ \bibinfo {pages} {4490} (\bibinfo {year} {2021})}\BibitemShut {NoStop}%
\bibitem [{\citenamefont {Ba{\~n}uls}\ \emph {et~al.}(2020)\citenamefont {Ba{\~n}uls}, \citenamefont {Blatt}, \citenamefont {Catani}, \citenamefont {Celi}, \citenamefont {Cirac}, \citenamefont {Dalmonte}, \citenamefont {Fallani}, \citenamefont {Jansen}, \citenamefont {Lewenstein}, \citenamefont {Montangero}, \citenamefont {Muschik}, \citenamefont {Reznik}, \citenamefont {Rico}, \citenamefont {Tagliacozzo}, \citenamefont {Van~Acoleyen}, \citenamefont {Verstraete}, \citenamefont {Wiese}, \citenamefont {Wingate}, \citenamefont {Zakrzewski},\ and\ \citenamefont {Zoller}}]{Banuls_2020}%
  \BibitemOpen
  \bibfield  {author} {\bibinfo {author} {\bibfnamefont {M.~C.}\ \bibnamefont {Ba{\~n}uls}}, \bibinfo {author} {\bibfnamefont {R.}~\bibnamefont {Blatt}}, \bibinfo {author} {\bibfnamefont {J.}~\bibnamefont {Catani}}, \bibinfo {author} {\bibfnamefont {A.}~\bibnamefont {Celi}}, \bibinfo {author} {\bibfnamefont {J.~I.}\ \bibnamefont {Cirac}}, \bibinfo {author} {\bibfnamefont {M.}~\bibnamefont {Dalmonte}}, \bibinfo {author} {\bibfnamefont {L.}~\bibnamefont {Fallani}}, \bibinfo {author} {\bibfnamefont {K.}~\bibnamefont {Jansen}}, \bibinfo {author} {\bibfnamefont {M.}~\bibnamefont {Lewenstein}}, \bibinfo {author} {\bibfnamefont {S.}~\bibnamefont {Montangero}}, \bibinfo {author} {\bibfnamefont {C.~A.}\ \bibnamefont {Muschik}}, \bibinfo {author} {\bibfnamefont {B.}~\bibnamefont {Reznik}}, \bibinfo {author} {\bibfnamefont {E.}~\bibnamefont {Rico}}, \bibinfo {author} {\bibfnamefont {L.}~\bibnamefont {Tagliacozzo}}, \bibinfo {author} {\bibfnamefont {K.}~\bibnamefont {Van~Acoleyen}}, \bibinfo {author} {\bibfnamefont
  {F.}~\bibnamefont {Verstraete}}, \bibinfo {author} {\bibfnamefont {U.-J.}\ \bibnamefont {Wiese}}, \bibinfo {author} {\bibfnamefont {M.}~\bibnamefont {Wingate}}, \bibinfo {author} {\bibfnamefont {J.}~\bibnamefont {Zakrzewski}},\ and\ \bibinfo {author} {\bibfnamefont {P.}~\bibnamefont {Zoller}},\ }\bibfield  {title} {\bibinfo {title} {Simulating lattice gauge theories within quantum technologies},\ }\href {https://doi.org/10.1140/epjd/e2020-100571-8} {\bibfield  {journal} {\bibinfo  {journal} {The European Physical Journal D}\ }\textbf {\bibinfo {volume} {74}},\ \bibinfo {pages} {165} (\bibinfo {year} {2020})}\BibitemShut {NoStop}%
\bibitem [{\citenamefont {Aidelsburger}\ \emph {et~al.}(2022)\citenamefont {Aidelsburger}, \citenamefont {Barbiero}, \citenamefont {Bermudez}, \citenamefont {Chanda}, \citenamefont {Dauphin}, \citenamefont {Gonz{\'{a}}lez-Cuadra}, \citenamefont {Grzybowski}, \citenamefont {Hands}, \citenamefont {Jendrzejewski}, \citenamefont {J{\"{u}}nemann}, \citenamefont {Juzeli{\={u}}nas}, \citenamefont {Kasper}, \citenamefont {Piga}, \citenamefont {Ran}, \citenamefont {Rizzi}, \citenamefont {Sierra}, \citenamefont {Tagliacozzo}, \citenamefont {Tirrito}, \citenamefont {Zache}, \citenamefont {Zakrzewski}, \citenamefont {Zohar},\ and\ \citenamefont {Lewenstein}}]{Aidelsburger2022}%
  \BibitemOpen
  \bibfield  {author} {\bibinfo {author} {\bibfnamefont {M.}~\bibnamefont {Aidelsburger}}, \bibinfo {author} {\bibfnamefont {L.}~\bibnamefont {Barbiero}}, \bibinfo {author} {\bibfnamefont {A.}~\bibnamefont {Bermudez}}, \bibinfo {author} {\bibfnamefont {T.}~\bibnamefont {Chanda}}, \bibinfo {author} {\bibfnamefont {A.}~\bibnamefont {Dauphin}}, \bibinfo {author} {\bibfnamefont {D.}~\bibnamefont {Gonz{\'{a}}lez-Cuadra}}, \bibinfo {author} {\bibfnamefont {P.~R.}\ \bibnamefont {Grzybowski}}, \bibinfo {author} {\bibfnamefont {S.}~\bibnamefont {Hands}}, \bibinfo {author} {\bibfnamefont {F.}~\bibnamefont {Jendrzejewski}}, \bibinfo {author} {\bibfnamefont {J.}~\bibnamefont {J{\"{u}}nemann}}, \bibinfo {author} {\bibfnamefont {G.}~\bibnamefont {Juzeli{\={u}}nas}}, \bibinfo {author} {\bibfnamefont {V.}~\bibnamefont {Kasper}}, \bibinfo {author} {\bibfnamefont {A.}~\bibnamefont {Piga}}, \bibinfo {author} {\bibfnamefont {S.~J.}\ \bibnamefont {Ran}}, \bibinfo {author} {\bibfnamefont {M.}~\bibnamefont {Rizzi}}, \bibinfo
  {author} {\bibfnamefont {G.}~\bibnamefont {Sierra}}, \bibinfo {author} {\bibfnamefont {L.}~\bibnamefont {Tagliacozzo}}, \bibinfo {author} {\bibfnamefont {E.}~\bibnamefont {Tirrito}}, \bibinfo {author} {\bibfnamefont {T.~V.}\ \bibnamefont {Zache}}, \bibinfo {author} {\bibfnamefont {J.}~\bibnamefont {Zakrzewski}}, \bibinfo {author} {\bibfnamefont {E.}~\bibnamefont {Zohar}},\ and\ \bibinfo {author} {\bibfnamefont {M.}~\bibnamefont {Lewenstein}},\ }\bibfield  {title} {\bibinfo {title} {{Cold atoms meet lattice gauge theory}},\ }\href {https://doi.org/10.1098/RSTA.2021.0064} {\bibfield  {journal} {\bibinfo  {journal} {Philosophical Transactions of the Royal Society A}\ }\textbf {\bibinfo {volume} {380}},\ \bibinfo {pages} {20210064} (\bibinfo {year} {2022})}\BibitemShut {NoStop}%
\bibitem [{\citenamefont {Di~Meglio}\ \emph {et~al.}(2024)\citenamefont {Di~Meglio}, \citenamefont {Jansen}, \citenamefont {Tavernelli}, \citenamefont {Alexandrou}, \citenamefont {Arunachalam}, \citenamefont {Bauer}, \citenamefont {Borras}, \citenamefont {Carrazza}, \citenamefont {Crippa}, \citenamefont {Croft}, \citenamefont {de~Putter}, \citenamefont {Delgado}, \citenamefont {Dunjko}, \citenamefont {Egger}, \citenamefont {Fern\'andez-Combarro}, \citenamefont {Fuchs}, \citenamefont {Funcke}, \citenamefont {Gonz\'alez-Cuadra}, \citenamefont {Grossi}, \citenamefont {Halimeh}, \citenamefont {Holmes}, \citenamefont {K\"uhn}, \citenamefont {Lacroix}, \citenamefont {Lewis}, \citenamefont {Lucchesi}, \citenamefont {Martinez}, \citenamefont {Meloni}, \citenamefont {Mezzacapo}, \citenamefont {Montangero}, \citenamefont {Nagano}, \citenamefont {Pascuzzi}, \citenamefont {Radescu}, \citenamefont {Ortega}, \citenamefont {Roggero}, \citenamefont {Schuhmacher}, \citenamefont {Seixas}, \citenamefont {Silvi}, \citenamefont
  {Spentzouris}, \citenamefont {Tacchino}, \citenamefont {Temme}, \citenamefont {Terashi}, \citenamefont {Tura}, \citenamefont {T\"uys\"uz}, \citenamefont {Vallecorsa}, \citenamefont {Wiese}, \citenamefont {Yoo},\ and\ \citenamefont {Zhang}}]{DiMeglio_2024}%
  \BibitemOpen
  \bibfield  {author} {\bibinfo {author} {\bibfnamefont {A.}~\bibnamefont {Di~Meglio}}, \bibinfo {author} {\bibfnamefont {K.}~\bibnamefont {Jansen}}, \bibinfo {author} {\bibfnamefont {I.}~\bibnamefont {Tavernelli}}, \bibinfo {author} {\bibfnamefont {C.}~\bibnamefont {Alexandrou}}, \bibinfo {author} {\bibfnamefont {S.}~\bibnamefont {Arunachalam}}, \bibinfo {author} {\bibfnamefont {C.~W.}\ \bibnamefont {Bauer}}, \bibinfo {author} {\bibfnamefont {K.}~\bibnamefont {Borras}}, \bibinfo {author} {\bibfnamefont {S.}~\bibnamefont {Carrazza}}, \bibinfo {author} {\bibfnamefont {A.}~\bibnamefont {Crippa}}, \bibinfo {author} {\bibfnamefont {V.}~\bibnamefont {Croft}}, \bibinfo {author} {\bibfnamefont {R.}~\bibnamefont {de~Putter}}, \bibinfo {author} {\bibfnamefont {A.}~\bibnamefont {Delgado}}, \bibinfo {author} {\bibfnamefont {V.}~\bibnamefont {Dunjko}}, \bibinfo {author} {\bibfnamefont {D.~J.}\ \bibnamefont {Egger}}, \bibinfo {author} {\bibfnamefont {E.}~\bibnamefont {Fern\'andez-Combarro}}, \bibinfo {author}
  {\bibfnamefont {E.}~\bibnamefont {Fuchs}}, \bibinfo {author} {\bibfnamefont {L.}~\bibnamefont {Funcke}}, \bibinfo {author} {\bibfnamefont {D.}~\bibnamefont {Gonz\'alez-Cuadra}}, \bibinfo {author} {\bibfnamefont {M.}~\bibnamefont {Grossi}}, \bibinfo {author} {\bibfnamefont {J.~C.}\ \bibnamefont {Halimeh}}, \bibinfo {author} {\bibfnamefont {Z.}~\bibnamefont {Holmes}}, \bibinfo {author} {\bibfnamefont {S.}~\bibnamefont {K\"uhn}}, \bibinfo {author} {\bibfnamefont {D.}~\bibnamefont {Lacroix}}, \bibinfo {author} {\bibfnamefont {R.}~\bibnamefont {Lewis}}, \bibinfo {author} {\bibfnamefont {D.}~\bibnamefont {Lucchesi}}, \bibinfo {author} {\bibfnamefont {M.~L.}\ \bibnamefont {Martinez}}, \bibinfo {author} {\bibfnamefont {F.}~\bibnamefont {Meloni}}, \bibinfo {author} {\bibfnamefont {A.}~\bibnamefont {Mezzacapo}}, \bibinfo {author} {\bibfnamefont {S.}~\bibnamefont {Montangero}}, \bibinfo {author} {\bibfnamefont {L.}~\bibnamefont {Nagano}}, \bibinfo {author} {\bibfnamefont {V.~R.}\ \bibnamefont {Pascuzzi}}, \bibinfo
  {author} {\bibfnamefont {V.}~\bibnamefont {Radescu}}, \bibinfo {author} {\bibfnamefont {E.~R.}\ \bibnamefont {Ortega}}, \bibinfo {author} {\bibfnamefont {A.}~\bibnamefont {Roggero}}, \bibinfo {author} {\bibfnamefont {J.}~\bibnamefont {Schuhmacher}}, \bibinfo {author} {\bibfnamefont {J.}~\bibnamefont {Seixas}}, \bibinfo {author} {\bibfnamefont {P.}~\bibnamefont {Silvi}}, \bibinfo {author} {\bibfnamefont {P.}~\bibnamefont {Spentzouris}}, \bibinfo {author} {\bibfnamefont {F.}~\bibnamefont {Tacchino}}, \bibinfo {author} {\bibfnamefont {K.}~\bibnamefont {Temme}}, \bibinfo {author} {\bibfnamefont {K.}~\bibnamefont {Terashi}}, \bibinfo {author} {\bibfnamefont {J.}~\bibnamefont {Tura}}, \bibinfo {author} {\bibfnamefont {C.}~\bibnamefont {T\"uys\"uz}}, \bibinfo {author} {\bibfnamefont {S.}~\bibnamefont {Vallecorsa}}, \bibinfo {author} {\bibfnamefont {U.-J.}\ \bibnamefont {Wiese}}, \bibinfo {author} {\bibfnamefont {S.}~\bibnamefont {Yoo}},\ and\ \bibinfo {author} {\bibfnamefont {J.}~\bibnamefont {Zhang}},\ }\bibfield
   {title} {\bibinfo {title} {Quantum computing for high-energy physics: State of the art and challenges},\ }\href {https://doi.org/10.1103/PRXQuantum.5.037001} {\bibfield  {journal} {\bibinfo  {journal} {PRX Quantum}\ }\textbf {\bibinfo {volume} {5}},\ \bibinfo {pages} {037001} (\bibinfo {year} {2024})}\BibitemShut {NoStop}%
\bibitem [{\citenamefont {{Arg{\"u}ello-Luengo}}\ \emph {et~al.}(2019)\citenamefont {{Arg{\"u}ello-Luengo}}, \citenamefont {{Gonz{\'a}lez-Tudela}}, \citenamefont {Shi}, \citenamefont {Zoller},\ and\ \citenamefont {Cirac}}]{arguello-luengoAnalogue2019}%
  \BibitemOpen
  \bibfield  {author} {\bibinfo {author} {\bibfnamefont {J.}~\bibnamefont {{Arg{\"u}ello-Luengo}}}, \bibinfo {author} {\bibfnamefont {A.}~\bibnamefont {{Gonz{\'a}lez-Tudela}}}, \bibinfo {author} {\bibfnamefont {T.}~\bibnamefont {Shi}}, \bibinfo {author} {\bibfnamefont {P.}~\bibnamefont {Zoller}},\ and\ \bibinfo {author} {\bibfnamefont {J.~I.}\ \bibnamefont {Cirac}},\ }\bibfield  {title} {\bibinfo {title} {Analogue quantum chemistry simulation},\ }\href {https://doi.org/10.1038/s41586-019-1614-4} {\bibfield  {journal} {\bibinfo  {journal} {Nature}\ }\textbf {\bibinfo {volume} {574}},\ \bibinfo {pages} {215} (\bibinfo {year} {2019})}\BibitemShut {NoStop}%
\bibitem [{\citenamefont {{Arg{\"u}ello-Luengo}}\ \emph {et~al.}(2020)\citenamefont {{Arg{\"u}ello-Luengo}}, \citenamefont {{Gonz{\'a}lez-Tudela}}, \citenamefont {Shi}, \citenamefont {Zoller},\ and\ \citenamefont {Cirac}}]{arguello-luengoQuantum2020}%
  \BibitemOpen
  \bibfield  {author} {\bibinfo {author} {\bibfnamefont {J.}~\bibnamefont {{Arg{\"u}ello-Luengo}}}, \bibinfo {author} {\bibfnamefont {A.}~\bibnamefont {{Gonz{\'a}lez-Tudela}}}, \bibinfo {author} {\bibfnamefont {T.}~\bibnamefont {Shi}}, \bibinfo {author} {\bibfnamefont {P.}~\bibnamefont {Zoller}},\ and\ \bibinfo {author} {\bibfnamefont {J.~I.}\ \bibnamefont {Cirac}},\ }\bibfield  {title} {\bibinfo {title} {Quantum simulation of two-dimensional quantum chemistry in optical lattices},\ }\href {https://doi.org/10.1103/PhysRevResearch.2.042013} {\bibfield  {journal} {\bibinfo  {journal} {Physical Review Research}\ }\textbf {\bibinfo {volume} {2}},\ \bibinfo {pages} {042013} (\bibinfo {year} {2020})}\BibitemShut {NoStop}%
\bibitem [{\citenamefont {{Arg{\"u}ello-Luengo}}\ \emph {et~al.}(2021)\citenamefont {{Arg{\"u}ello-Luengo}}, \citenamefont {Shi},\ and\ \citenamefont {{Gonz{\'a}lez-Tudela}}}]{arguello-luengoEngineering2021}%
  \BibitemOpen
  \bibfield  {author} {\bibinfo {author} {\bibfnamefont {J.}~\bibnamefont {{Arg{\"u}ello-Luengo}}}, \bibinfo {author} {\bibfnamefont {T.}~\bibnamefont {Shi}},\ and\ \bibinfo {author} {\bibfnamefont {A.}~\bibnamefont {{Gonz{\'a}lez-Tudela}}},\ }\bibfield  {title} {\bibinfo {title} {Engineering analog quantum chemistry {{Hamiltonians}} using cold atoms in optical lattices},\ }\href {https://doi.org/10.1103/PhysRevA.103.043318} {\bibfield  {journal} {\bibinfo  {journal} {Phys. Rev. A}\ }\textbf {\bibinfo {volume} {103}},\ \bibinfo {pages} {043318} (\bibinfo {year} {2021})}\BibitemShut {NoStop}%
\bibitem [{\citenamefont {Malz}\ and\ \citenamefont {Cirac}(2023)}]{malz2023}%
  \BibitemOpen
  \bibfield  {author} {\bibinfo {author} {\bibfnamefont {D.}~\bibnamefont {Malz}}\ and\ \bibinfo {author} {\bibfnamefont {J.~I.}\ \bibnamefont {Cirac}},\ }\bibfield  {title} {\bibinfo {title} {{Few-Body Analog Quantum Simulation with Rydberg-Dressed Atoms in Optical Lattices}},\ }\href {https://doi.org/10.1103/PRXQUANTUM.4.020301/FIGURES/4/MEDIUM} {\bibfield  {journal} {\bibinfo  {journal} {PRX Quantum}\ }\textbf {\bibinfo {volume} {4}},\ \bibinfo {pages} {020301} (\bibinfo {year} {2023})}\BibitemShut {NoStop}%
\bibitem [{\citenamefont {MacDonell}\ \emph {et~al.}(2021)\citenamefont {MacDonell}, \citenamefont {Dickerson}, \citenamefont {Birch}, \citenamefont {Kumar}, \citenamefont {Edmunds}, \citenamefont {Biercuk}, \citenamefont {Hempel},\ and\ \citenamefont {Kassal}}]{MaDonnel2021}%
  \BibitemOpen
  \bibfield  {author} {\bibinfo {author} {\bibfnamefont {R.~J.}\ \bibnamefont {MacDonell}}, \bibinfo {author} {\bibfnamefont {C.~E.}\ \bibnamefont {Dickerson}}, \bibinfo {author} {\bibfnamefont {C.~J.}\ \bibnamefont {Birch}}, \bibinfo {author} {\bibfnamefont {A.}~\bibnamefont {Kumar}}, \bibinfo {author} {\bibfnamefont {C.~L.}\ \bibnamefont {Edmunds}}, \bibinfo {author} {\bibfnamefont {M.~J.}\ \bibnamefont {Biercuk}}, \bibinfo {author} {\bibfnamefont {C.}~\bibnamefont {Hempel}},\ and\ \bibinfo {author} {\bibfnamefont {I.}~\bibnamefont {Kassal}},\ }\bibfield  {title} {\bibinfo {title} {{Analog quantum simulation of chemical dynamics}},\ }\href {https://doi.org/10.1039/D1SC02142G} {\bibfield  {journal} {\bibinfo  {journal} {Chemical Science}\ }\textbf {\bibinfo {volume} {12}},\ \bibinfo {pages} {9794} (\bibinfo {year} {2021})}\BibitemShut {NoStop}%
\bibitem [{\citenamefont {Navickas}\ \emph {et~al.}(2024)\citenamefont {Navickas}, \citenamefont {MacDonell}, \citenamefont {Valahu}, \citenamefont {Olaya-Agudelo}, \citenamefont {Scuccimarra}, \citenamefont {Millican}, \citenamefont {Matsos}, \citenamefont {Nourse}, \citenamefont {Rao}, \citenamefont {Biercuk}, \citenamefont {Hempel}, \citenamefont {Kassal},\ and\ \citenamefont {Tan}}]{Navickas2024}%
  \BibitemOpen
  \bibfield  {author} {\bibinfo {author} {\bibfnamefont {T.}~\bibnamefont {Navickas}}, \bibinfo {author} {\bibfnamefont {R.~J.}\ \bibnamefont {MacDonell}}, \bibinfo {author} {\bibfnamefont {C.~H.}\ \bibnamefont {Valahu}}, \bibinfo {author} {\bibfnamefont {V.~C.}\ \bibnamefont {Olaya-Agudelo}}, \bibinfo {author} {\bibfnamefont {F.}~\bibnamefont {Scuccimarra}}, \bibinfo {author} {\bibfnamefont {M.~J.}\ \bibnamefont {Millican}}, \bibinfo {author} {\bibfnamefont {V.~G.}\ \bibnamefont {Matsos}}, \bibinfo {author} {\bibfnamefont {H.~L.}\ \bibnamefont {Nourse}}, \bibinfo {author} {\bibfnamefont {A.~D.}\ \bibnamefont {Rao}}, \bibinfo {author} {\bibfnamefont {M.~J.}\ \bibnamefont {Biercuk}}, \bibinfo {author} {\bibfnamefont {C.}~\bibnamefont {Hempel}}, \bibinfo {author} {\bibfnamefont {I.}~\bibnamefont {Kassal}},\ and\ \bibinfo {author} {\bibfnamefont {T.~R.}\ \bibnamefont {Tan}},\ }\bibfield  {title} {\bibinfo {title} {{Experimental Quantum Simulation of Chemical Dynamics}},\ }\href
  {https://doi.org/10.1021/jacs.5c03336} {\bibfield  {journal} {\bibinfo  {journal} {J. Am. Chem. Soc.}\ }\textbf {\bibinfo {volume} {147}},\ \bibinfo {pages} {23566–23573} (\bibinfo {year} {2025})}\BibitemShut {NoStop}%
\bibitem [{\citenamefont {Valahu}\ \emph {et~al.}(2023)\citenamefont {Valahu}, \citenamefont {Olaya-Agudelo}, \citenamefont {MacDonell}, \citenamefont {Navickas}, \citenamefont {Rao}, \citenamefont {Millican}, \citenamefont {P{\'{e}}rez-S{\'{a}}nchez}, \citenamefont {Yuen-Zhou}, \citenamefont {Biercuk}, \citenamefont {Hempel}, \citenamefont {Tan},\ and\ \citenamefont {Kassal}}]{Valahu2023}%
  \BibitemOpen
  \bibfield  {author} {\bibinfo {author} {\bibfnamefont {C.~H.}\ \bibnamefont {Valahu}}, \bibinfo {author} {\bibfnamefont {V.~C.}\ \bibnamefont {Olaya-Agudelo}}, \bibinfo {author} {\bibfnamefont {R.~J.}\ \bibnamefont {MacDonell}}, \bibinfo {author} {\bibfnamefont {T.}~\bibnamefont {Navickas}}, \bibinfo {author} {\bibfnamefont {A.~D.}\ \bibnamefont {Rao}}, \bibinfo {author} {\bibfnamefont {M.~J.}\ \bibnamefont {Millican}}, \bibinfo {author} {\bibfnamefont {J.~B.}\ \bibnamefont {P{\'{e}}rez-S{\'{a}}nchez}}, \bibinfo {author} {\bibfnamefont {J.}~\bibnamefont {Yuen-Zhou}}, \bibinfo {author} {\bibfnamefont {M.~J.}\ \bibnamefont {Biercuk}}, \bibinfo {author} {\bibfnamefont {C.}~\bibnamefont {Hempel}}, \bibinfo {author} {\bibfnamefont {T.~R.}\ \bibnamefont {Tan}},\ and\ \bibinfo {author} {\bibfnamefont {I.}~\bibnamefont {Kassal}},\ }\bibfield  {title} {\bibinfo {title} {{Direct observation of geometric-phase interference in dynamics around a conical intersection}},\ }\href
  {https://doi.org/10.1038/s41557-023-01300-3} {\bibfield  {journal} {\bibinfo  {journal} {Nature Chemistry}\ }\textbf {\bibinfo {volume} {15}},\ \bibinfo {pages} {1503} (\bibinfo {year} {2023})}\BibitemShut {NoStop}%
\bibitem [{\citenamefont {Olaya-Agudelo}\ \emph {et~al.}(2024)\citenamefont {Olaya-Agudelo}, \citenamefont {Stewart}, \citenamefont {Valahu}, \citenamefont {MacDonell}, \citenamefont {Millican}, \citenamefont {Matsos}, \citenamefont {Scuccimarra}, \citenamefont {Tan},\ and\ \citenamefont {Kassal}}]{Olaya2024}%
  \BibitemOpen
  \bibfield  {author} {\bibinfo {author} {\bibfnamefont {V.~C.}\ \bibnamefont {Olaya-Agudelo}}, \bibinfo {author} {\bibfnamefont {B.}~\bibnamefont {Stewart}}, \bibinfo {author} {\bibfnamefont {C.~H.}\ \bibnamefont {Valahu}}, \bibinfo {author} {\bibfnamefont {R.~J.}\ \bibnamefont {MacDonell}}, \bibinfo {author} {\bibfnamefont {M.~J.}\ \bibnamefont {Millican}}, \bibinfo {author} {\bibfnamefont {V.~G.}\ \bibnamefont {Matsos}}, \bibinfo {author} {\bibfnamefont {F.}~\bibnamefont {Scuccimarra}}, \bibinfo {author} {\bibfnamefont {T.~R.}\ \bibnamefont {Tan}},\ and\ \bibinfo {author} {\bibfnamefont {I.}~\bibnamefont {Kassal}},\ }\bibfield  {title} {\bibinfo {title} {{Simulating open-system molecular dynamics on analog quantum computers}},\ }\href {DOI: https://doi.org/10.1103/PhysRevResearch.7.023215} {\bibfield  {journal} {\bibinfo  {journal} {Phys. Rev. Research}\ }\textbf {\bibinfo {volume} {7}},\ \bibinfo {pages} {023215} (\bibinfo {year} {2025})}\BibitemShut {NoStop}%
\bibitem [{\citenamefont {Ha}\ and\ \citenamefont {MacDonell}(2024)}]{haAnalog2024}%
  \BibitemOpen
  \bibfield  {author} {\bibinfo {author} {\bibfnamefont {J.-K.}\ \bibnamefont {Ha}}\ and\ \bibinfo {author} {\bibfnamefont {R.~J.}\ \bibnamefont {MacDonell}},\ }\href {https://doi.org/10.48550/arXiv.2409.04427} {\bibinfo {title} {Analog {{Quantum Simulation}} of {{Coupled Electron-Nuclear Dynamics}} in {{Molecules}}}} (\bibinfo {year} {2024}),\ \Eprint {https://arxiv.org/abs/2409.04427} {arXiv:2409.04427 [physics, physics:quant-ph]} \BibitemShut {NoStop}%
\bibitem [{\citenamefont {Szabo}\ and\ \citenamefont {Ostlund}(2012)}]{szabo2012modern}%
  \BibitemOpen
  \bibfield  {author} {\bibinfo {author} {\bibfnamefont {A.}~\bibnamefont {Szabo}}\ and\ \bibinfo {author} {\bibfnamefont {N.~S.}\ \bibnamefont {Ostlund}},\ }\href@noop {} {\emph {\bibinfo {title} {Modern Quantum Chemistry: Introduction to Advanced Electronic Structure Theory}}}\ (\bibinfo  {publisher} {Courier Corporation},\ \bibinfo {year} {2012})\BibitemShut {NoStop}%
\bibitem [{\citenamefont {Christakis}\ \emph {et~al.}(2023)\citenamefont {Christakis}, \citenamefont {Rosenberg}, \citenamefont {Raj}, \citenamefont {Chi}, \citenamefont {Morningstar}, \citenamefont {Huse}, \citenamefont {Yan},\ and\ \citenamefont {Bakr}}]{christakisProbing2023}%
  \BibitemOpen
  \bibfield  {author} {\bibinfo {author} {\bibfnamefont {L.}~\bibnamefont {Christakis}}, \bibinfo {author} {\bibfnamefont {J.~S.}\ \bibnamefont {Rosenberg}}, \bibinfo {author} {\bibfnamefont {R.}~\bibnamefont {Raj}}, \bibinfo {author} {\bibfnamefont {S.}~\bibnamefont {Chi}}, \bibinfo {author} {\bibfnamefont {A.}~\bibnamefont {Morningstar}}, \bibinfo {author} {\bibfnamefont {D.~A.}\ \bibnamefont {Huse}}, \bibinfo {author} {\bibfnamefont {Z.~Z.}\ \bibnamefont {Yan}},\ and\ \bibinfo {author} {\bibfnamefont {W.~S.}\ \bibnamefont {Bakr}},\ }\bibfield  {title} {\bibinfo {title} {Probing site-resolved correlations in a spin system of ultracold molecules},\ }\href {https://doi.org/10.1038/s41586-022-05558-4} {\bibfield  {journal} {\bibinfo  {journal} {Nature}\ }\textbf {\bibinfo {volume} {614}},\ \bibinfo {pages} {64} (\bibinfo {year} {2023})}\BibitemShut {NoStop}%
\bibitem [{\citenamefont {Yan}\ \emph {et~al.}(2013)\citenamefont {Yan}, \citenamefont {Moses}, \citenamefont {Gadway}, \citenamefont {Covey}, \citenamefont {Hazzard}, \citenamefont {Rey}, \citenamefont {Jin},\ and\ \citenamefont {Ye}}]{yan13a}%
  \BibitemOpen
  \bibfield  {author} {\bibinfo {author} {\bibfnamefont {B.}~\bibnamefont {Yan}}, \bibinfo {author} {\bibfnamefont {S.~A.}\ \bibnamefont {Moses}}, \bibinfo {author} {\bibfnamefont {B.}~\bibnamefont {Gadway}}, \bibinfo {author} {\bibfnamefont {J.~P.}\ \bibnamefont {Covey}}, \bibinfo {author} {\bibfnamefont {K.~R.~A.}\ \bibnamefont {Hazzard}}, \bibinfo {author} {\bibfnamefont {A.~M.}\ \bibnamefont {Rey}}, \bibinfo {author} {\bibfnamefont {D.~S.}\ \bibnamefont {Jin}},\ and\ \bibinfo {author} {\bibfnamefont {J.}~\bibnamefont {Ye}},\ }\bibfield  {title} {\bibinfo {title} {{Observation of dipolar spin-exchange interactions with lattice-confined polar molecules}},\ }\href@noop {} {\bibfield  {journal} {\bibinfo  {journal} {Nature}\ }\textbf {\bibinfo {volume} {501}},\ \bibinfo {pages} {521} (\bibinfo {year} {2013})}\BibitemShut {NoStop}%
\bibitem [{\citenamefont {Chotia}\ \emph {et~al.}(2012)\citenamefont {Chotia}, \citenamefont {Neyenhuis}, \citenamefont {Moses}, \citenamefont {Yan}, \citenamefont {Covey}, \citenamefont {{Foss-Feig}}, \citenamefont {Rey}, \citenamefont {Jin},\ and\ \citenamefont {Ye}}]{chotiaLongLived2012}%
  \BibitemOpen
  \bibfield  {author} {\bibinfo {author} {\bibfnamefont {A.}~\bibnamefont {Chotia}}, \bibinfo {author} {\bibfnamefont {B.}~\bibnamefont {Neyenhuis}}, \bibinfo {author} {\bibfnamefont {S.~A.}\ \bibnamefont {Moses}}, \bibinfo {author} {\bibfnamefont {B.}~\bibnamefont {Yan}}, \bibinfo {author} {\bibfnamefont {J.~P.}\ \bibnamefont {Covey}}, \bibinfo {author} {\bibfnamefont {M.}~\bibnamefont {{Foss-Feig}}}, \bibinfo {author} {\bibfnamefont {A.~M.}\ \bibnamefont {Rey}}, \bibinfo {author} {\bibfnamefont {D.~S.}\ \bibnamefont {Jin}},\ and\ \bibinfo {author} {\bibfnamefont {J.}~\bibnamefont {Ye}},\ }\bibfield  {title} {\bibinfo {title} {Long-{{Lived Dipolar Molecules}} and {{Feshbach Molecules}} in a {{3D Optical Lattice}}},\ }\href {https://doi.org/10.1103/PhysRevLett.108.080405} {\bibfield  {journal} {\bibinfo  {journal} {Phys. Rev. Lett.}\ }\textbf {\bibinfo {volume} {108}},\ \bibinfo {pages} {080405} (\bibinfo {year} {2012})}\BibitemShut {NoStop}%
\bibitem [{\citenamefont {Brown}\ and\ \citenamefont {Carrington}(2003)}]{brownRotational2003}%
  \BibitemOpen
  \bibfield  {author} {\bibinfo {author} {\bibfnamefont {J.~M.}\ \bibnamefont {Brown}}\ and\ \bibinfo {author} {\bibfnamefont {A.}~\bibnamefont {Carrington}},\ }\href@noop {} {\emph {\bibinfo {title} {Rotational {{Spectroscopy}} of {{Diatomic Molecules}}}}}\ (\bibinfo  {publisher} {Cambridge University Press},\ \bibinfo {year} {2003})\BibitemShut {NoStop}%
\bibitem [{\citenamefont {Wall}\ \emph {et~al.}(2015)\citenamefont {Wall}, \citenamefont {Hazzard},\ and\ \citenamefont {Rey}}]{wall2015Quantum}%
  \BibitemOpen
  \bibfield  {author} {\bibinfo {author} {\bibfnamefont {M.}~\bibnamefont {Wall}}, \bibinfo {author} {\bibfnamefont {K.}~\bibnamefont {Hazzard}},\ and\ \bibinfo {author} {\bibfnamefont {A.~M.}\ \bibnamefont {Rey}},\ }\bibfield  {title} {\bibinfo {title} {Quantum magnetism with ultracold molecules},\ }in\ \href@noop {} {\emph {\bibinfo {booktitle} {From atomic to mesoscale: the role of quantum coherence in systems of various complexities}}}\ (\bibinfo  {publisher} {World Scientific},\ \bibinfo {year} {2015})\ pp.\ \bibinfo {pages} {3--37}\BibitemShut {NoStop}%
\bibitem [{\citenamefont {Micheli}\ \emph {et~al.}(2006)\citenamefont {Micheli}, \citenamefont {Brennen},\ and\ \citenamefont {Zoller}}]{micheliToolbox2006}%
  \BibitemOpen
  \bibfield  {author} {\bibinfo {author} {\bibfnamefont {A.}~\bibnamefont {Micheli}}, \bibinfo {author} {\bibfnamefont {G.~K.}\ \bibnamefont {Brennen}},\ and\ \bibinfo {author} {\bibfnamefont {P.}~\bibnamefont {Zoller}},\ }\bibfield  {title} {\bibinfo {title} {A toolbox for lattice-spin models with polar molecules},\ }\href {https://doi.org/10.1038/nphys287} {\bibfield  {journal} {\bibinfo  {journal} {Nature Phys}\ }\textbf {\bibinfo {volume} {2}},\ \bibinfo {pages} {341} (\bibinfo {year} {2006})}\BibitemShut {NoStop}%
\bibitem [{\citenamefont {Gorshkov}\ \emph {et~al.}(2011)\citenamefont {Gorshkov}, \citenamefont {Manmana}, \citenamefont {Chen}, \citenamefont {Ye}, \citenamefont {Demler}, \citenamefont {Lukin},\ and\ \citenamefont {Rey}}]{gorshkovTunable2011}%
  \BibitemOpen
  \bibfield  {author} {\bibinfo {author} {\bibfnamefont {A.~V.}\ \bibnamefont {Gorshkov}}, \bibinfo {author} {\bibfnamefont {S.~R.}\ \bibnamefont {Manmana}}, \bibinfo {author} {\bibfnamefont {G.}~\bibnamefont {Chen}}, \bibinfo {author} {\bibfnamefont {J.}~\bibnamefont {Ye}}, \bibinfo {author} {\bibfnamefont {E.}~\bibnamefont {Demler}}, \bibinfo {author} {\bibfnamefont {M.~D.}\ \bibnamefont {Lukin}},\ and\ \bibinfo {author} {\bibfnamefont {A.~M.}\ \bibnamefont {Rey}},\ }\bibfield  {title} {\bibinfo {title} {Tunable {{Superfluidity}} and {{Quantum Magnetism}} with {{Ultracold Polar Molecules}}},\ }\href {https://doi.org/10.1103/PhysRevLett.107.115301} {\bibfield  {journal} {\bibinfo  {journal} {Phys. Rev. Lett.}\ }\textbf {\bibinfo {volume} {107}},\ \bibinfo {pages} {115301} (\bibinfo {year} {2011})}\BibitemShut {NoStop}%
\bibitem [{\citenamefont {Micheli}\ \emph {et~al.}(2007)\citenamefont {Micheli}, \citenamefont {Pupillo}, \citenamefont {B{\"u}chler},\ and\ \citenamefont {Zoller}}]{micheliCold2007}%
  \BibitemOpen
  \bibfield  {author} {\bibinfo {author} {\bibfnamefont {A.}~\bibnamefont {Micheli}}, \bibinfo {author} {\bibfnamefont {G.}~\bibnamefont {Pupillo}}, \bibinfo {author} {\bibfnamefont {H.~P.}\ \bibnamefont {B{\"u}chler}},\ and\ \bibinfo {author} {\bibfnamefont {P.}~\bibnamefont {Zoller}},\ }\bibfield  {title} {\bibinfo {title} {Cold polar molecules in two-dimensional traps: {{Tailoring}} interactions with external fields for novel quantum phases},\ }\href {https://doi.org/10.1103/PhysRevA.76.043604} {\bibfield  {journal} {\bibinfo  {journal} {Phys. Rev. A}\ }\textbf {\bibinfo {volume} {76}},\ \bibinfo {pages} {043604} (\bibinfo {year} {2007})}\BibitemShut {NoStop}%
\bibitem [{\citenamefont {Wall}(2015)}]{wallHyperfine2015}%
  \BibitemOpen
  \bibfield  {author} {\bibinfo {author} {\bibfnamefont {M.~L.}\ \bibnamefont {Wall}},\ }\bibfield  {title} {\bibinfo {title} {Hyperfine {{Molecular Hubbard Hamiltonian}}},\ }in\ \href {https://doi.org/10.1007/978-3-319-14252-4_4} {\emph {\bibinfo {booktitle} {Quantum {{Many-Body Physics}} of {{Ultracold Molecules}} in {{Optical Lattices}}: {{Models}} and {{Simulation Methods}}}}},\ \bibinfo {editor} {edited by\ \bibinfo {editor} {\bibfnamefont {M.~L.}\ \bibnamefont {Wall}}}\ (\bibinfo  {publisher} {Springer International Publishing},\ \bibinfo {address} {Cham},\ \bibinfo {year} {2015})\ pp.\ \bibinfo {pages} {93--119}\BibitemShut {NoStop}%
\bibitem [{\citenamefont {Yan}\ \emph {et~al.}(2020)\citenamefont {Yan}, \citenamefont {Park}, \citenamefont {Ni}, \citenamefont {Loh}, \citenamefont {Will}, \citenamefont {Karman},\ and\ \citenamefont {Zwierlein}}]{yanResonant2020}%
  \BibitemOpen
  \bibfield  {author} {\bibinfo {author} {\bibfnamefont {Z.~Z.}\ \bibnamefont {Yan}}, \bibinfo {author} {\bibfnamefont {J.~W.}\ \bibnamefont {Park}}, \bibinfo {author} {\bibfnamefont {Y.}~\bibnamefont {Ni}}, \bibinfo {author} {\bibfnamefont {H.}~\bibnamefont {Loh}}, \bibinfo {author} {\bibfnamefont {S.}~\bibnamefont {Will}}, \bibinfo {author} {\bibfnamefont {T.}~\bibnamefont {Karman}},\ and\ \bibinfo {author} {\bibfnamefont {M.}~\bibnamefont {Zwierlein}},\ }\bibfield  {title} {\bibinfo {title} {Resonant {{Dipolar Collisions}} of {{Ultracold Molecules Induced}} by {{Microwave Dressing}}},\ }\href {https://doi.org/10.1103/PhysRevLett.125.063401} {\bibfield  {journal} {\bibinfo  {journal} {Phys. Rev. Lett.}\ }\textbf {\bibinfo {volume} {125}},\ \bibinfo {pages} {063401} (\bibinfo {year} {2020})}\BibitemShut {NoStop}%
\bibitem [{EM()}]{EM}%
  \BibitemOpen
  \href@noop {} {}\bibinfo {howpublished} {See the End Matter acccompanying this Letter for further details about the dipolar interaction, the engineering of the wavepacket and the numerical benchmark of the proposal.}\BibitemShut {Stop}%
\bibitem [{Note1()}]{Note1}%
  \BibitemOpen
  \bibinfo {note} {While the effective Hamiltonian holds valid for inter-site interactions, dipolar interactions will dominate over the electric field for separations $r<r_d\equiv (d^2/B_N)^3 \sim 10$ nm. As $r_d<a$, this will only affect on-site interactions, which motivates our choice of effective potential $V_\protect \text {mol}$~\cite {yanResonant2020,janssenCold2012}. Microwave shielding to repulsive resonant dipolar interactions could potentially be used to prevent molecules from getting closer to each other~\cite {karmanResonant2022,yanResonant2020}.}\BibitemShut {Stop}%
\bibitem [{\citenamefont {Lombardi}\ and\ \citenamefont {Palazzetti}(2018)}]{lombardiChirality2018}%
  \BibitemOpen
  \bibfield  {author} {\bibinfo {author} {\bibfnamefont {A.}~\bibnamefont {Lombardi}}\ and\ \bibinfo {author} {\bibfnamefont {F.}~\bibnamefont {Palazzetti}},\ }\bibfield  {title} {\bibinfo {title} {Chirality in molecular collision dynamics},\ }\href {https://doi.org/10.1088/1361-648X/aaa1c8} {\bibfield  {journal} {\bibinfo  {journal} {J. Phys.: Condens. Matter}\ }\textbf {\bibinfo {volume} {30}},\ \bibinfo {pages} {063003} (\bibinfo {year} {2018})}\BibitemShut {NoStop}%
\bibitem [{\citenamefont {Domcke}\ \emph {et~al.}(2004)\citenamefont {Domcke}, \citenamefont {Koppel},\ and\ \citenamefont {Yarkony}}]{domckeConical2004}%
  \BibitemOpen
  \bibfield  {author} {\bibinfo {author} {\bibfnamefont {W.}~\bibnamefont {Domcke}}, \bibinfo {author} {\bibfnamefont {H.}~\bibnamefont {Koppel}},\ and\ \bibinfo {author} {\bibfnamefont {D.~R.}\ \bibnamefont {Yarkony}},\ }\href@noop {} {\emph {\bibinfo {title} {Conical {{Intersections}}: {{Electronic Structure}}, {{Dynamics}} \& {{Spectroscopy}}}}}\ (\bibinfo  {publisher} {World Scientific},\ \bibinfo {year} {2004})\BibitemShut {NoStop}%
\bibitem [{\citenamefont {Mead}\ and\ \citenamefont {Truhlar}(1979)}]{meadDetermination1979}%
  \BibitemOpen
  \bibfield  {author} {\bibinfo {author} {\bibfnamefont {C.~A.}\ \bibnamefont {Mead}}\ and\ \bibinfo {author} {\bibfnamefont {D.~G.}\ \bibnamefont {Truhlar}},\ }\bibfield  {title} {\bibinfo {title} {On the determination of {{Born}}--{{Oppenheimer}} nuclear motion wave functions including complications due to conical intersections and identical nuclei},\ }\href {https://doi.org/10.1063/1.437734} {\bibfield  {journal} {\bibinfo  {journal} {The Journal of Chemical Physics}\ }\textbf {\bibinfo {volume} {70}},\ \bibinfo {pages} {2284} (\bibinfo {year} {1979})}\BibitemShut {NoStop}%
\bibitem [{\citenamefont {Wu}\ and\ \citenamefont {Kuppermann}(1993)}]{wuPrediction1993}%
  \BibitemOpen
  \bibfield  {author} {\bibinfo {author} {\bibfnamefont {Y.-S.~M.}\ \bibnamefont {Wu}}\ and\ \bibinfo {author} {\bibfnamefont {A.}~\bibnamefont {Kuppermann}},\ }\bibfield  {title} {\bibinfo {title} {Prediction of the effect of the geometric phase on product rotational state distributions and integral cross sections},\ }\href {https://doi.org/10.1016/0009-2614(93)85053-Q} {\bibfield  {journal} {\bibinfo  {journal} {Chemical Physics Letters}\ }\textbf {\bibinfo {volume} {201}},\ \bibinfo {pages} {178} (\bibinfo {year} {1993})}\BibitemShut {NoStop}%
\bibitem [{\citenamefont {Althorpe}\ \emph {et~al.}(2002)\citenamefont {Althorpe}, \citenamefont {{Fern{\'a}ndez-Alonso}}, \citenamefont {Bean}, \citenamefont {Ayers}, \citenamefont {Pomerantz}, \citenamefont {Zare},\ and\ \citenamefont {Wrede}}]{althorpeObservation2002}%
  \BibitemOpen
  \bibfield  {author} {\bibinfo {author} {\bibfnamefont {S.~C.}\ \bibnamefont {Althorpe}}, \bibinfo {author} {\bibfnamefont {F.}~\bibnamefont {{Fern{\'a}ndez-Alonso}}}, \bibinfo {author} {\bibfnamefont {B.~D.}\ \bibnamefont {Bean}}, \bibinfo {author} {\bibfnamefont {J.~D.}\ \bibnamefont {Ayers}}, \bibinfo {author} {\bibfnamefont {A.~E.}\ \bibnamefont {Pomerantz}}, \bibinfo {author} {\bibfnamefont {R.~N.}\ \bibnamefont {Zare}},\ and\ \bibinfo {author} {\bibfnamefont {E.}~\bibnamefont {Wrede}},\ }\bibfield  {title} {\bibinfo {title} {Observation and interpretation of a time-delayed mechanism in the hydrogen exchange reaction},\ }\href {https://doi.org/10.1038/416067a} {\bibfield  {journal} {\bibinfo  {journal} {Nature}\ }\textbf {\bibinfo {volume} {416}},\ \bibinfo {pages} {67} (\bibinfo {year} {2002})}\BibitemShut {NoStop}%
\bibitem [{\citenamefont {{Rubio-Abadal}}\ \emph {et~al.}(2019)\citenamefont {{Rubio-Abadal}}, \citenamefont {Choi}, \citenamefont {Zeiher}, \citenamefont {Hollerith}, \citenamefont {Rui}, \citenamefont {Bloch},\ and\ \citenamefont {Gross}}]{rubio-abadalManyBody2019}%
  \BibitemOpen
  \bibfield  {author} {\bibinfo {author} {\bibfnamefont {A.}~\bibnamefont {{Rubio-Abadal}}}, \bibinfo {author} {\bibfnamefont {J.-y.}\ \bibnamefont {Choi}}, \bibinfo {author} {\bibfnamefont {J.}~\bibnamefont {Zeiher}}, \bibinfo {author} {\bibfnamefont {S.}~\bibnamefont {Hollerith}}, \bibinfo {author} {\bibfnamefont {J.}~\bibnamefont {Rui}}, \bibinfo {author} {\bibfnamefont {I.}~\bibnamefont {Bloch}},\ and\ \bibinfo {author} {\bibfnamefont {C.}~\bibnamefont {Gross}},\ }\bibfield  {title} {\bibinfo {title} {Many-{{Body Delocalization}} in the {{Presence}} of a {{Quantum Bath}}},\ }\href {https://doi.org/10.1103/PhysRevX.9.041014} {\bibfield  {journal} {\bibinfo  {journal} {Phys. Rev. X}\ }\textbf {\bibinfo {volume} {9}},\ \bibinfo {pages} {041014} (\bibinfo {year} {2019})}\BibitemShut {NoStop}%
\bibitem [{\citenamefont {Fukuhara}\ \emph {et~al.}(2013)\citenamefont {Fukuhara}, \citenamefont {Kantian}, \citenamefont {Endres}, \citenamefont {Cheneau}, \citenamefont {Schau{\ss}}, \citenamefont {Hild}, \citenamefont {Bellem}, \citenamefont {Schollw{\"o}ck}, \citenamefont {Giamarchi}, \citenamefont {Gross}, \citenamefont {Bloch},\ and\ \citenamefont {Kuhr}}]{fukuharaQuantum2013}%
  \BibitemOpen
  \bibfield  {author} {\bibinfo {author} {\bibfnamefont {T.}~\bibnamefont {Fukuhara}}, \bibinfo {author} {\bibfnamefont {A.}~\bibnamefont {Kantian}}, \bibinfo {author} {\bibfnamefont {M.}~\bibnamefont {Endres}}, \bibinfo {author} {\bibfnamefont {M.}~\bibnamefont {Cheneau}}, \bibinfo {author} {\bibfnamefont {P.}~\bibnamefont {Schau{\ss}}}, \bibinfo {author} {\bibfnamefont {S.}~\bibnamefont {Hild}}, \bibinfo {author} {\bibfnamefont {D.}~\bibnamefont {Bellem}}, \bibinfo {author} {\bibfnamefont {U.}~\bibnamefont {Schollw{\"o}ck}}, \bibinfo {author} {\bibfnamefont {T.}~\bibnamefont {Giamarchi}}, \bibinfo {author} {\bibfnamefont {C.}~\bibnamefont {Gross}}, \bibinfo {author} {\bibfnamefont {I.}~\bibnamefont {Bloch}},\ and\ \bibinfo {author} {\bibfnamefont {S.}~\bibnamefont {Kuhr}},\ }\bibfield  {title} {\bibinfo {title} {Quantum dynamics of a mobile spin impurity},\ }\href {https://doi.org/10.1038/nphys2561} {\bibfield  {journal} {\bibinfo  {journal} {Nature Phys}\ }\textbf {\bibinfo {volume} {9}},\ \bibinfo {pages}
  {235} (\bibinfo {year} {2013})}\BibitemShut {NoStop}%
\bibitem [{\citenamefont {Karski}\ \emph {et~al.}(2009)\citenamefont {Karski}, \citenamefont {F{\"o}rster}, \citenamefont {Choi}, \citenamefont {Steffen}, \citenamefont {Alt}, \citenamefont {Meschede},\ and\ \citenamefont {Widera}}]{karskiQuantum2009}%
  \BibitemOpen
  \bibfield  {author} {\bibinfo {author} {\bibfnamefont {M.}~\bibnamefont {Karski}}, \bibinfo {author} {\bibfnamefont {L.}~\bibnamefont {F{\"o}rster}}, \bibinfo {author} {\bibfnamefont {J.-M.}\ \bibnamefont {Choi}}, \bibinfo {author} {\bibfnamefont {A.}~\bibnamefont {Steffen}}, \bibinfo {author} {\bibfnamefont {W.}~\bibnamefont {Alt}}, \bibinfo {author} {\bibfnamefont {D.}~\bibnamefont {Meschede}},\ and\ \bibinfo {author} {\bibfnamefont {A.}~\bibnamefont {Widera}},\ }\bibfield  {title} {\bibinfo {title} {Quantum {{Walk}} in {{Position Space}} with {{Single Optically Trapped Atoms}}},\ }\href {https://doi.org/10.1126/science.1174436} {\bibfield  {journal} {\bibinfo  {journal} {Science}\ }\textbf {\bibinfo {volume} {325}},\ \bibinfo {pages} {174} (\bibinfo {year} {2009})}\BibitemShut {NoStop}%
\bibitem [{\citenamefont {Young}\ \emph {et~al.}(2022)\citenamefont {Young}, \citenamefont {Eckner}, \citenamefont {Schine}, \citenamefont {Childs},\ and\ \citenamefont {Kaufman}}]{youngTweezerprogrammable2022}%
  \BibitemOpen
  \bibfield  {author} {\bibinfo {author} {\bibfnamefont {A.~W.}\ \bibnamefont {Young}}, \bibinfo {author} {\bibfnamefont {W.~J.}\ \bibnamefont {Eckner}}, \bibinfo {author} {\bibfnamefont {N.}~\bibnamefont {Schine}}, \bibinfo {author} {\bibfnamefont {A.~M.}\ \bibnamefont {Childs}},\ and\ \bibinfo {author} {\bibfnamefont {A.~M.}\ \bibnamefont {Kaufman}},\ }\bibfield  {title} {\bibinfo {title} {Tweezer-programmable {{2D}} quantum walks in a {{Hubbard-regime}} lattice},\ }\href {https://doi.org/10.1126/science.abo0608} {\bibfield  {journal} {\bibinfo  {journal} {Science}\ }\textbf {\bibinfo {volume} {377}},\ \bibinfo {pages} {885} (\bibinfo {year} {2022})}\BibitemShut {NoStop}%
\bibitem [{\citenamefont {Lauber}\ \emph {et~al.}(2011)\citenamefont {Lauber}, \citenamefont {Massignan}, \citenamefont {Birkl},\ and\ \citenamefont {Sanpera}}]{lauberAtomic2011}%
  \BibitemOpen
  \bibfield  {author} {\bibinfo {author} {\bibfnamefont {T.}~\bibnamefont {Lauber}}, \bibinfo {author} {\bibfnamefont {P.}~\bibnamefont {Massignan}}, \bibinfo {author} {\bibfnamefont {G.}~\bibnamefont {Birkl}},\ and\ \bibinfo {author} {\bibfnamefont {A.}~\bibnamefont {Sanpera}},\ }\bibfield  {title} {\bibinfo {title} {Atomic wave packet dynamics in finite time-dependent optical lattices},\ }\href {https://doi.org/10.1088/0953-4075/44/6/065301} {\bibfield  {journal} {\bibinfo  {journal} {J. Phys. B: At. Mol. Opt. Phys.}\ }\textbf {\bibinfo {volume} {44}},\ \bibinfo {pages} {065301} (\bibinfo {year} {2011})}\BibitemShut {NoStop}%
\bibitem [{\citenamefont {Fabre}\ \emph {et~al.}(2011)\citenamefont {Fabre}, \citenamefont {Cheiney}, \citenamefont {Gattobigio}, \citenamefont {Vermersch}, \citenamefont {Faure}, \citenamefont {Mathevet}, \citenamefont {Lahaye},\ and\ \citenamefont {{Gu{\'e}ry-Odelin}}}]{fabreRealization2011}%
  \BibitemOpen
  \bibfield  {author} {\bibinfo {author} {\bibfnamefont {C.~M.}\ \bibnamefont {Fabre}}, \bibinfo {author} {\bibfnamefont {P.}~\bibnamefont {Cheiney}}, \bibinfo {author} {\bibfnamefont {G.~L.}\ \bibnamefont {Gattobigio}}, \bibinfo {author} {\bibfnamefont {F.}~\bibnamefont {Vermersch}}, \bibinfo {author} {\bibfnamefont {S.}~\bibnamefont {Faure}}, \bibinfo {author} {\bibfnamefont {R.}~\bibnamefont {Mathevet}}, \bibinfo {author} {\bibfnamefont {T.}~\bibnamefont {Lahaye}},\ and\ \bibinfo {author} {\bibfnamefont {D.}~\bibnamefont {{Gu{\'e}ry-Odelin}}},\ }\bibfield  {title} {\bibinfo {title} {Realization of a {{Distributed Bragg Reflector}} for {{Propagating Guided Matter Waves}}},\ }\href {https://doi.org/10.1103/PhysRevLett.107.230401} {\bibfield  {journal} {\bibinfo  {journal} {Phys. Rev. Lett.}\ }\textbf {\bibinfo {volume} {107}},\ \bibinfo {pages} {230401} (\bibinfo {year} {2011})}\BibitemShut {NoStop}%
\bibitem [{\citenamefont {Su}\ \emph {et~al.}(2024)\citenamefont {Su}, \citenamefont {Osborne},\ and\ \citenamefont {Halimeh}}]{suColdAtom2024}%
  \BibitemOpen
  \bibfield  {author} {\bibinfo {author} {\bibfnamefont {G.-X.}\ \bibnamefont {Su}}, \bibinfo {author} {\bibfnamefont {J.}~\bibnamefont {Osborne}},\ and\ \bibinfo {author} {\bibfnamefont {J.~C.}\ \bibnamefont {Halimeh}},\ }\href {https://doi.org/10.48550/arXiv.2401.05489} {\bibinfo {title} {Cold-Atom Particle Collide}, \ } \href {https://doi.org/10.1103/PRXQuantum.5.040310} {\bibfield  {journal} {\bibinfo  {journal} {PRX Quantum}\ }\textbf {\bibinfo {volume} {5}},\ \bibinfo {pages} {040310} (\bibinfo {year} {2024})}\BibitemShut {NoStop}%
\bibitem [{\citenamefont {Bause}\ \emph {et~al.}(2020)\citenamefont {Bause}, \citenamefont {Li}, \citenamefont {Schindewolf}, \citenamefont {Chen}, \citenamefont {Duda}, \citenamefont {Kotochigova}, \citenamefont {Bloch},\ and\ \citenamefont {Luo}}]{bauseTuneOut2020}%
  \BibitemOpen
  \bibfield  {author} {\bibinfo {author} {\bibfnamefont {R.}~\bibnamefont {Bause}}, \bibinfo {author} {\bibfnamefont {M.}~\bibnamefont {Li}}, \bibinfo {author} {\bibfnamefont {A.}~\bibnamefont {Schindewolf}}, \bibinfo {author} {\bibfnamefont {X.-Y.}\ \bibnamefont {Chen}}, \bibinfo {author} {\bibfnamefont {M.}~\bibnamefont {Duda}}, \bibinfo {author} {\bibfnamefont {S.}~\bibnamefont {Kotochigova}}, \bibinfo {author} {\bibfnamefont {I.}~\bibnamefont {Bloch}},\ and\ \bibinfo {author} {\bibfnamefont {X.-Y.}\ \bibnamefont {Luo}},\ }\bibfield  {title} {\bibinfo {title} {Tune-{Out} and {Magic} {Wavelengths} for {Ground}-{State} {23Na40K} {Molecules}},\ }\href {https://doi.org/10.1103/PhysRevLett.125.023201} {\bibfield  {journal} {\bibinfo  {journal} {Physical Review Letters}\ }\textbf {\bibinfo {volume} {125}},\ \bibinfo {pages} {023201} (\bibinfo {year} {2020})}\BibitemShut {NoStop}%
\bibitem [{\citenamefont {Ye}\ \emph {et~al.}(2008)\citenamefont {Ye}, \citenamefont {Kimble},\ and\ \citenamefont {Katori}}]{yeQuantum2008}%
  \BibitemOpen
  \bibfield  {author} {\bibinfo {author} {\bibfnamefont {J.}~\bibnamefont {Ye}}, \bibinfo {author} {\bibfnamefont {H.~J.}\ \bibnamefont {Kimble}},\ and\ \bibinfo {author} {\bibfnamefont {H.}~\bibnamefont {Katori}},\ }\bibfield  {title} {\bibinfo {title} {Quantum {State} {Engineering} and {Precision} {Metrology} {Using} {State}-{Insensitive} {Light} {Traps}},\ }\href {https://doi.org/10.1126/science.1148259} {\bibfield  {journal} {\bibinfo  {journal} {Science}\ }\textbf {\bibinfo {volume} {320}},\ \bibinfo {pages} {1734} (\bibinfo {year} {2008})}\BibitemShut {NoStop}%
\bibitem [{\citenamefont {Guan}\ \emph {et~al.}(2021)\citenamefont {Guan}, \citenamefont {Cornish},\ and\ \citenamefont {Kotochigova}}]{guanMagic2021}%
  \BibitemOpen
  \bibfield  {author} {\bibinfo {author} {\bibfnamefont {Q.}~\bibnamefont {Guan}}, \bibinfo {author} {\bibfnamefont {S.~L.}\ \bibnamefont {Cornish}},\ and\ \bibinfo {author} {\bibfnamefont {S.}~\bibnamefont {Kotochigova}},\ }\bibfield  {title} {\bibinfo {title} {Magic conditions for multiple rotational states of bialkali molecules in optical lattices},\ }\href {https://doi.org/10.1103/PhysRevA.103.043311} {\bibfield  {journal} {\bibinfo  {journal} {Physical Review A}\ }\textbf {\bibinfo {volume} {103}},\ \bibinfo {pages} {043311} (\bibinfo {year} {2021})}\BibitemShut {NoStop}%
\bibitem [{\citenamefont {Fite}\ and\ \citenamefont {Brackmann}(1958)}]{fiteCollisions1958}%
  \BibitemOpen
  \bibfield  {author} {\bibinfo {author} {\bibfnamefont {W.~L.}\ \bibnamefont {Fite}}\ and\ \bibinfo {author} {\bibfnamefont {R.~T.}\ \bibnamefont {Brackmann}},\ }\bibfield  {title} {\bibinfo {title} {Collisions of {{Electrons}} with {{Hydrogen Atoms}}. {{I}}. {{Ionization}}},\ }\href {https://doi.org/10.1103/PhysRev.112.1141} {\bibfield  {journal} {\bibinfo  {journal} {Phys. Rev.}\ }\textbf {\bibinfo {volume} {112}},\ \bibinfo {pages} {1141} (\bibinfo {year} {1958})}\BibitemShut {NoStop}%
\bibitem [{\citenamefont {Rudge}(1968)}]{rudgeTheory1968}%
  \BibitemOpen
  \bibfield  {author} {\bibinfo {author} {\bibfnamefont {M.~R.~H.}\ \bibnamefont {Rudge}},\ }\bibfield  {title} {\bibinfo {title} {Theory of the {{Ionization}} of {{Atoms}} by {{Electron Impact}}},\ }\href {https://doi.org/10.1103/RevModPhys.40.564} {\bibfield  {journal} {\bibinfo  {journal} {Rev. Mod. Phys.}\ }\textbf {\bibinfo {volume} {40}},\ \bibinfo {pages} {564} (\bibinfo {year} {1968})}\BibitemShut {NoStop}%
\bibitem [{\citenamefont {Bransden}\ \emph {et~al.}(2003)\citenamefont {Bransden}, \citenamefont {Joachain},\ and\ \citenamefont {Plivier}}]{bransden2003physics}%
  \BibitemOpen
  \bibfield  {author} {\bibinfo {author} {\bibfnamefont {B.~H.}\ \bibnamefont {Bransden}}, \bibinfo {author} {\bibfnamefont {C.~J.}\ \bibnamefont {Joachain}},\ and\ \bibinfo {author} {\bibfnamefont {T.~J.}\ \bibnamefont {Plivier}},\ }\href@noop {} {\emph {\bibinfo {title} {Physics of Atoms and Molecules}}}\ (\bibinfo  {publisher} {Pearson education},\ \bibinfo {year} {2003})\BibitemShut {NoStop}%
\bibitem [{\citenamefont {Gross}\ and\ \citenamefont {Bakr}(2021{\natexlab{b}})}]{Gross2021}%
  \BibitemOpen
  \bibfield  {author} {\bibinfo {author} {\bibfnamefont {C.}~\bibnamefont {Gross}}\ and\ \bibinfo {author} {\bibfnamefont {W.~S.}\ \bibnamefont {Bakr}},\ }\bibfield  {title} {\bibinfo {title} {{Quantum gas microscopy for single atom and spin detection}},\ }\bibfield  {journal} {\bibinfo  {journal} {Nat. Phys.}\ }\textbf {\bibinfo {volume} {17}},\ \href {https://doi.org/10.1038/s41567-021-01370-5} {10.1038/s41567-021-01370-5} (\bibinfo {year} {2021}{\natexlab{b}})\BibitemShut {NoStop}%
\bibitem [{\citenamefont {Weitenberg}\ \emph {et~al.}(2011)\citenamefont {Weitenberg}, \citenamefont {Endres}, \citenamefont {Sherson}, \citenamefont {Cheneau}, \citenamefont {Schau{\ss}}, \citenamefont {Fukuhara}, \citenamefont {Bloch},\ and\ \citenamefont {Kuhr}}]{Weitenberg2011}%
  \BibitemOpen
  \bibfield  {author} {\bibinfo {author} {\bibfnamefont {C.}~\bibnamefont {Weitenberg}}, \bibinfo {author} {\bibfnamefont {M.}~\bibnamefont {Endres}}, \bibinfo {author} {\bibfnamefont {J.~F.}\ \bibnamefont {Sherson}}, \bibinfo {author} {\bibfnamefont {M.}~\bibnamefont {Cheneau}}, \bibinfo {author} {\bibfnamefont {P.}~\bibnamefont {Schau{\ss}}}, \bibinfo {author} {\bibfnamefont {T.}~\bibnamefont {Fukuhara}}, \bibinfo {author} {\bibfnamefont {I.}~\bibnamefont {Bloch}},\ and\ \bibinfo {author} {\bibfnamefont {S.}~\bibnamefont {Kuhr}},\ }\bibfield  {title} {\bibinfo {title} {Single-spin addressing in an atomic {{Mott}} insulator},\ }\href {https://doi.org/10.1038/nature09827} {\bibfield  {journal} {\bibinfo  {journal} {Nature}\ }\textbf {\bibinfo {volume} {471}},\ \bibinfo {pages} {319} (\bibinfo {year} {2011})}\BibitemShut {NoStop}%
\bibitem [{\citenamefont {Bennewitz}\ \emph {et~al.}(2024)\citenamefont {Bennewitz}, \citenamefont {Ware}, \citenamefont {Schuckert}, \citenamefont {Lerose}, \citenamefont {Surace}, \citenamefont {Belyansky}, \citenamefont {Morong}, \citenamefont {Luo}, \citenamefont {De}, \citenamefont {Collins}, \citenamefont {Katz}, \citenamefont {Monroe}, \citenamefont {Davoudi},\ and\ \citenamefont {Gorshkov}}]{bennewitzSimulating2024}%
  \BibitemOpen
  \bibfield  {author} {\bibinfo {author} {\bibfnamefont {E.~R.}\ \bibnamefont {Bennewitz}}, \bibinfo {author} {\bibfnamefont {B.}~\bibnamefont {Ware}}, \bibinfo {author} {\bibfnamefont {A.}~\bibnamefont {Schuckert}}, \bibinfo {author} {\bibfnamefont {A.}~\bibnamefont {Lerose}}, \bibinfo {author} {\bibfnamefont {F.~M.}\ \bibnamefont {Surace}}, \bibinfo {author} {\bibfnamefont {R.}~\bibnamefont {Belyansky}}, \bibinfo {author} {\bibfnamefont {W.}~\bibnamefont {Morong}}, \bibinfo {author} {\bibfnamefont {D.}~\bibnamefont {Luo}}, \bibinfo {author} {\bibfnamefont {A.}~\bibnamefont {De}}, \bibinfo {author} {\bibfnamefont {K.~S.}\ \bibnamefont {Collins}}, \bibinfo {author} {\bibfnamefont {O.}~\bibnamefont {Katz}}, \bibinfo {author} {\bibfnamefont {C.}~\bibnamefont {Monroe}}, \bibinfo {author} {\bibfnamefont {Z.}~\bibnamefont {Davoudi}},\ and\ \bibinfo {author} {\bibfnamefont {A.~V.}\ \bibnamefont {Gorshkov}},\ }\href {	https://doi.org/10.22331/q-2025-06-17-1773} {\bibfield  {journal} {\bibinfo  {journal} {Quantum}\ }\textbf {\bibinfo {volume} {9}},\ \bibinfo {pages} {1773} (\bibinfo {year} {2025})}\BibitemShut {NoStop}%
\bibitem [{\citenamefont {Sharma}\ and\ \citenamefont {Hazzard}(2025)}]{sharmaMeson2025}%
  \BibitemOpen
  \bibfield  {author} {\bibinfo {author} {\bibfnamefont {V.}~\bibnamefont {Sharma}}\ and\ \bibinfo {author} {\bibfnamefont {K.~R.~A.}\ \bibnamefont {Hazzard}},\ }\href {https://doi.org/10.48550/arXiv.2503.02791} {\bibinfo {title} {Meson dynamics from locally exciting a particle-conserving \${{Z}}\_2\$ lattice gauge theory}} (\bibinfo {year} {2025}),\ \Eprint {https://arxiv.org/abs/2503.02791} {arXiv:2503.02791 [quant-ph]} \BibitemShut {NoStop}%
\bibitem [{\citenamefont {Norcia}\ \emph {et~al.}(2021)\citenamefont {Norcia}, \citenamefont {Politi}, \citenamefont {Klaus}, \citenamefont {Poli}, \citenamefont {Sohmen}, \citenamefont {Mark}, \citenamefont {Bisset}, \citenamefont {Santos},\ and\ \citenamefont {Ferlaino}}]{norciaTwodimensional2021}%
  \BibitemOpen
  \bibfield  {author} {\bibinfo {author} {\bibfnamefont {M.~A.}\ \bibnamefont {Norcia}}, \bibinfo {author} {\bibfnamefont {C.}~\bibnamefont {Politi}}, \bibinfo {author} {\bibfnamefont {L.}~\bibnamefont {Klaus}}, \bibinfo {author} {\bibfnamefont {E.}~\bibnamefont {Poli}}, \bibinfo {author} {\bibfnamefont {M.}~\bibnamefont {Sohmen}}, \bibinfo {author} {\bibfnamefont {M.~J.}\ \bibnamefont {Mark}}, \bibinfo {author} {\bibfnamefont {R.~N.}\ \bibnamefont {Bisset}}, \bibinfo {author} {\bibfnamefont {L.}~\bibnamefont {Santos}},\ and\ \bibinfo {author} {\bibfnamefont {F.}~\bibnamefont {Ferlaino}},\ }\bibfield  {title} {\bibinfo {title} {Two-dimensional supersolidity in a dipolar quantum gas},\ }\href {https://doi.org/10.1038/s41586-021-03725-7} {\bibfield  {journal} {\bibinfo  {journal} {Nature}\ }\textbf {\bibinfo {volume} {596}},\ \bibinfo {pages} {357} (\bibinfo {year} {2021})}\BibitemShut {NoStop}%
\bibitem [{\citenamefont {Baier}\ \emph {et~al.}(2016)\citenamefont {Baier}, \citenamefont {Mark}, \citenamefont {Petter}, \citenamefont {Aikawa}, \citenamefont {Chomaz}, \citenamefont {Cai}, \citenamefont {Baranov}, \citenamefont {Zoller},\ and\ \citenamefont {Ferlaino}}]{baierExtended2016}%
  \BibitemOpen
  \bibfield  {author} {\bibinfo {author} {\bibfnamefont {S.}~\bibnamefont {Baier}}, \bibinfo {author} {\bibfnamefont {M.~J.}\ \bibnamefont {Mark}}, \bibinfo {author} {\bibfnamefont {D.}~\bibnamefont {Petter}}, \bibinfo {author} {\bibfnamefont {K.}~\bibnamefont {Aikawa}}, \bibinfo {author} {\bibfnamefont {L.}~\bibnamefont {Chomaz}}, \bibinfo {author} {\bibfnamefont {Z.}~\bibnamefont {Cai}}, \bibinfo {author} {\bibfnamefont {M.}~\bibnamefont {Baranov}}, \bibinfo {author} {\bibfnamefont {P.}~\bibnamefont {Zoller}},\ and\ \bibinfo {author} {\bibfnamefont {F.}~\bibnamefont {Ferlaino}},\ }\bibfield  {title} {\bibinfo {title} {Extended {{Bose-Hubbard}} models with ultracold magnetic atoms},\ }\href {https://doi.org/10.1126/science.aac9812} {\bibfield  {journal} {\bibinfo  {journal} {Science}\ }\textbf {\bibinfo {volume} {352}},\ \bibinfo {pages} {201} (\bibinfo {year} {2016})}\BibitemShut {NoStop}%
\bibitem [{\citenamefont {Chomaz}\ \emph {et~al.}(2022)\citenamefont {Chomaz}, \citenamefont {{Ferrier-Barbut}}, \citenamefont {Ferlaino}, \citenamefont {{Laburthe-Tolra}}, \citenamefont {Lev},\ and\ \citenamefont {Pfau}}]{chomazDipolar2022}%
  \BibitemOpen
  \bibfield  {author} {\bibinfo {author} {\bibfnamefont {L.}~\bibnamefont {Chomaz}}, \bibinfo {author} {\bibfnamefont {I.}~\bibnamefont {{Ferrier-Barbut}}}, \bibinfo {author} {\bibfnamefont {F.}~\bibnamefont {Ferlaino}}, \bibinfo {author} {\bibfnamefont {B.}~\bibnamefont {{Laburthe-Tolra}}}, \bibinfo {author} {\bibfnamefont {B.~L.}\ \bibnamefont {Lev}},\ and\ \bibinfo {author} {\bibfnamefont {T.}~\bibnamefont {Pfau}},\ }\bibfield  {title} {\bibinfo {title} {Dipolar physics: A review of experiments with magnetic quantum gases},\ }\href {https://doi.org/10.1088/1361-6633/aca814} {\bibfield  {journal} {\bibinfo  {journal} {Rep. Prog. Phys.}\ }\textbf {\bibinfo {volume} {86}},\ \bibinfo {pages} {026401} (\bibinfo {year} {2022})}\BibitemShut {NoStop}%
\bibitem [{\citenamefont {Balewski}\ \emph {et~al.}(2014)\citenamefont {Balewski}, \citenamefont {Krupp}, \citenamefont {Gaj}, \citenamefont {Hofferberth}, \citenamefont {L{\"o}w},\ and\ \citenamefont {Pfau}}]{balewskiRydberg2014}%
  \BibitemOpen
  \bibfield  {author} {\bibinfo {author} {\bibfnamefont {J.~B.}\ \bibnamefont {Balewski}}, \bibinfo {author} {\bibfnamefont {A.~T.}\ \bibnamefont {Krupp}}, \bibinfo {author} {\bibfnamefont {A.}~\bibnamefont {Gaj}}, \bibinfo {author} {\bibfnamefont {S.}~\bibnamefont {Hofferberth}}, \bibinfo {author} {\bibfnamefont {R.}~\bibnamefont {L{\"o}w}},\ and\ \bibinfo {author} {\bibfnamefont {T.}~\bibnamefont {Pfau}},\ }\bibfield  {title} {\bibinfo {title} {Rydberg dressing: Understanding of collective many-body effects and implications for experiments},\ }\href {https://doi.org/10.1088/1367-2630/16/6/063012} {\bibfield  {journal} {\bibinfo  {journal} {New J. Phys.}\ }\textbf {\bibinfo {volume} {16}},\ \bibinfo {pages} {063012} (\bibinfo {year} {2014})}\BibitemShut {NoStop}%
\bibitem [{\citenamefont {{Guardado-Sanchez}}\ \emph {et~al.}(2021)\citenamefont {{Guardado-Sanchez}}, \citenamefont {Spar}, \citenamefont {Schauss}, \citenamefont {Belyansky}, \citenamefont {Young}, \citenamefont {Bienias}, \citenamefont {Gorshkov}, \citenamefont {Iadecola},\ and\ \citenamefont {Bakr}}]{guardado-sanchezQuench2021}%
  \BibitemOpen
  \bibfield  {author} {\bibinfo {author} {\bibfnamefont {E.}~\bibnamefont {{Guardado-Sanchez}}}, \bibinfo {author} {\bibfnamefont {B.~M.}\ \bibnamefont {Spar}}, \bibinfo {author} {\bibfnamefont {P.}~\bibnamefont {Schauss}}, \bibinfo {author} {\bibfnamefont {R.}~\bibnamefont {Belyansky}}, \bibinfo {author} {\bibfnamefont {J.~T.}\ \bibnamefont {Young}}, \bibinfo {author} {\bibfnamefont {P.}~\bibnamefont {Bienias}}, \bibinfo {author} {\bibfnamefont {A.~V.}\ \bibnamefont {Gorshkov}}, \bibinfo {author} {\bibfnamefont {T.}~\bibnamefont {Iadecola}},\ and\ \bibinfo {author} {\bibfnamefont {W.~S.}\ \bibnamefont {Bakr}},\ }\bibfield  {title} {\bibinfo {title} {Quench {{Dynamics}} of a {{Fermi Gas}} with {{Strong Nonlocal Interactions}}},\ }\href {https://doi.org/10.1103/PhysRevX.11.021036} {\bibfield  {journal} {\bibinfo  {journal} {Phys. Rev. X}\ }\textbf {\bibinfo {volume} {11}},\ \bibinfo {pages} {021036} (\bibinfo {year} {2021})}\BibitemShut {NoStop}%
\bibitem [{\citenamefont {Janssen}(2012)}]{janssenCold2012}%
  \BibitemOpen
  \bibfield  {author} {\bibinfo {author} {\bibfnamefont {L.~M.~C.}\ \bibnamefont {Janssen}},\ }\emph {\bibinfo {title} {Cold Collision Dynamics of {{NH}} Radicals}},\ \href@noop {} {Ph.D. thesis},\ \bibinfo  {school} {Radboud Universiteit Nijmegen} (\bibinfo {year} {2012})\BibitemShut {NoStop}%
\bibitem [{\citenamefont {Karman}\ \emph {et~al.}(2022)\citenamefont {Karman}, \citenamefont {Yan},\ and\ \citenamefont {Zwierlein}}]{karmanResonant2022}%
  \BibitemOpen
  \bibfield  {author} {\bibinfo {author} {\bibfnamefont {T.}~\bibnamefont {Karman}}, \bibinfo {author} {\bibfnamefont {Z.~Z.}\ \bibnamefont {Yan}},\ and\ \bibinfo {author} {\bibfnamefont {M.}~\bibnamefont {Zwierlein}},\ }\bibfield  {title} {\bibinfo {title} {Resonant and first-order dipolar interactions between ultracold 1S molecules in static and microwave electric fields},\ }\href {https://doi.org/10.1103/PhysRevA.105.013321} {\bibfield  {journal} {\bibinfo  {journal} {Phys. Rev. A}\ }\textbf {\bibinfo {volume} {105}},\ \bibinfo {pages} {013321} (\bibinfo {year} {2022})}\BibitemShut {NoStop}%
\bibitem [{\citenamefont {Schwartz}(2016)}]{schwartz2016lecture}%
  \BibitemOpen
  \bibfield  {author} {\bibinfo {author} {\bibfnamefont {M.}~\bibnamefont {Schwartz}},\ }\bibfield  {title} {\bibinfo {title} {Lecture 11: {{Wavepackets}} and dispersion},\ }\href@noop {} {\bibfield  {journal} {\bibinfo  {journal} {Cambridge, MA: Harvard University Press}\ }\textbf {\bibinfo {volume} {39}},\ \bibinfo {pages} {1} (\bibinfo {year} {2016})}\BibitemShut {NoStop}%
\bibitem [{\citenamefont {Lloyd}(1996)}]{Lloyd1996}%
  \BibitemOpen
  \bibfield  {author} {\bibinfo {author} {\bibfnamefont {S.}~\bibnamefont {Lloyd}},\ }\bibfield  {title} {\bibinfo {title} {Universal quantum simulators},\ }\href {https://doi.org/10.1126/science.273.5278.1073} {\bibfield  {journal} {\bibinfo  {journal} {Science}\ }\textbf {\bibinfo {volume} {273}},\ \bibinfo {pages} {1073} (\bibinfo {year} {1996})}\BibitemShut {NoStop}%
\bibitem [{\citenamefont {Childs}\ \emph {et~al.}(2021)\citenamefont {Childs}, \citenamefont {Su}, \citenamefont {Tran}, \citenamefont {Wiebe},\ and\ \citenamefont {Zhu}}]{childsTheory2021}%
  \BibitemOpen
  \bibfield  {author} {\bibinfo {author} {\bibfnamefont {A.~M.}\ \bibnamefont {Childs}}, \bibinfo {author} {\bibfnamefont {Y.}~\bibnamefont {Su}}, \bibinfo {author} {\bibfnamefont {M.~C.}\ \bibnamefont {Tran}}, \bibinfo {author} {\bibfnamefont {N.}~\bibnamefont {Wiebe}},\ and\ \bibinfo {author} {\bibfnamefont {S.}~\bibnamefont {Zhu}},\ }\bibfield  {title} {\bibinfo {title} {Theory of {{Trotter Error}} with {{Commutator Scaling}}},\ }\href {https://doi.org/10.1103/PhysRevX.11.011020} {\bibfield  {journal} {\bibinfo  {journal} {Phys. Rev. X}\ }\textbf {\bibinfo {volume} {11}},\ \bibinfo {pages} {011020} (\bibinfo {year} {2021})}\BibitemShut {NoStop}%
\end{thebibliography}
\end{document}